\RequirePackage{fix-cm}
\documentclass[smallextended]{svjour3}       %
\smartqed  %
\usepackage{graphicx}
\usepackage{amssymb}
\usepackage{url}
\usepackage{xcolor}
\usepackage{enumitem}
\usepackage{booktabs}
\usepackage{longtable}
\usepackage{geometry}
\usepackage{booktabs}
\usepackage[normalem]{ulem}
\useunder{\uline}{\ul}{}

\begin{document}

\title{An Audit of Machine Learning Experiments on Software Defect Prediction
}

\author{Giuseppe Destefanis\textsuperscript{1} \and     
        Leila Yousefi\textsuperscript{2} \and
        Martin Shepperd\textsuperscript{2}\thanks{Corresponding author: martin.shepperd@brunel.ac.uk} \and
        Allan Tucker\textsuperscript{2} \and
        Stephen Swift\textsuperscript{2} \and
        Steve Counsell\textsuperscript{2} \and
        Mahir Arzoky\textsuperscript{2}
}
\institute{
    \textsuperscript{1}Department of Computer Science, University College London, UK \\
    \textsuperscript{2}Department of Computer Science, Brunel University London, UB8 3PH, UK
}

\date{Received: date / Accepted: date}

\maketitle

\begin{abstract}
\noindent
    \textbf{Background:} Machine learning algorithms are increasingly being proposed to solve the problem of predicting defect-prone software components.  In this literature, computational experiments are the primary means of evaluating and comparing learners and the credibility of findings depends critically on their experimental design and reporting.\\
    \textbf{Objective:} This paper audits recent software defect prediction (SDP) experiments by assessing their experimental design, analysis and reporting practices against widely accepted norms from statistics, machine learning and empirical software engineering. Our aim is to characterise the current state of practice and evaluate the reproducibility of published findings.\\
    \textbf{Method:} We undertook an audit of relevant studies published from the SCOPUS database (2019-2023) focusing on their experimental design and analysis choices e.g., the outcome variables such as F-measure and the type of out of sample (OOS) validation regime, e.g., cross-validation, plus the statistical analysis and inference mechanisms. In all, we evaluated nine different study issues. This was complemented by an assessment of reproducibility using the instrument proposed by Gonz{\'a}lez-Barahona and Robles.\\
    \textbf{Results:} Our search located approximately 1,585 experiments in SDP (2019-2023), a substantial body of work. From this, we randomly sampled 101 ($ \approx 6.4\%$) papers, 61 journal and 40 conference papers.  Almost 50\% are behind `paywalls'.  We found considerable divergence in research practice.  The number of datasets used ranged 1-365, the number of learners or learner variants evaluated from 1-34 and the number of performance metrics from 1 to 9. Approximately 45\% of papers made use of formal statistical inference.  We detected a total of 427 issues distributed across 101 papers (median=4) with only one paper being entirely issue-free.  In terms of reproducibility, experiments ranged from near perfect to lacking almost all required information.  We also found two examples of tortured phrases and potential ``paper mill'' activity.\\
    \textbf{Conclusions:} Approaches to designing and reporting computational experiments varied greatly, but almost half the studies provided insufficient information such that reproduction would be challenging.  Overall, our audit suggests that as a research community, we have considerable scope for improvement.  Fortunately, many improvements should be neither difficult nor costly to achieve. 

\end{abstract}

\section{Introduction}\label{Sec:Intro}

\begin{quote}
    ``A variety of recent studies, primarily in the biomedical field, have revealed that an uncomfortably large number of research results found in the literature fail this [quality] test, because of sloppy experimental methods, flawed statistical analyses, or in rare cases, fraud.''  ACM Artifact Review and Badging (2020) \cite{ACM2020}.\\
\end{quote}

\noindent
Sadly, there is mounting evidence that software engineering is not immune from such quality problems of questionable research methods and poor reporting.  This is compounded by growing activities of ``paper mills'' and other sources of fake papers \cite{Candal2022retracted}.  While it might seem somewhat negative to focus on problems, we argue that this is an essential step in the journey to improve research practice.  

Although machine learning algorithms are being widely touted as an effective means of classifying software components into those that are likely to be defect-prone and those that are not, it is proving difficult to obtain an overall picture of what this large and rapidly growing body of research is actually telling us. For example, in 2012 Menzies and Shepperd published an editorial on the lack of ``conclusion stability'' \cite{Menzies2012} in the field. A subsequent meta-analysis indicated that who conducted the research, i.e., the team, was a major source of variability in results \cite{Shepperd2014}. Similarly, Li et al.~\cite{Li2019} found that ``predictive power is heavily influenced by the evaluation metrics and testing procedure''.  

Baltes and Ralph~\cite{Baltes2022} undertook a more general review of empirical software engineering studies and exposed widespread problems with the way samples are constructed and interpreted. More specifically to software defect prediction (SDP), Liu et al.~\cite{Liu2021} in their systematic review of deep learning algorithms, showed cause for considerable concern regarding both the reproducibility and the replicability of such research.  Related questions are being asked about the underlying research quality, experimental design and reporting of much of this growing body of work, e.g., \cite{Shepperd2019,Liem2020}.  These concerns are likely contributors to the low take up by practitioners \cite{Rana2014,Stradowski2022}. 

So why is this audit called for?  First, SDP is an extremely active and growing research area.  Second, combining these many studies into a coherent body of knowledge is proving quite challenging, \cite{Mohammadi2023bayesian}.  Third, much of the methodological research literature focuses on experiments using human participants so we thought it would be interesting to seek some contrasts with computational experiments.

This paper makes the following contributions in that we:
\begin{enumerate}
    \item conduct an in depth audit (i.e., a systematic technical review against established methodological standards) of computational experiments in SDP over the past five years (2019-2023), 
    \item examine factors associated with (i) study reproducibility, (ii) experimental and reporting issues\footnote{We prefer the less emotive term of `issue' to `problem' because we do not wish this audit to be construed as some kind of ``witch hunt''.},
    \item make a set of recommendations as to how we as an SDP research community might improve our practice. 
\end{enumerate}

Please note that in the spirit of reproducibility, our raw data, the meta-data and analysis code (R embedded in an RMarkdown notebook) can be found in our Zenodo repository \cite{destefanis_2024_13927602}.

The remainder of the paper is organised as follows. In the next section, we review approaches to scrutiny and audit in science and particularly software engineering. In Section~\ref{Sec:Method}, we describe the conduct of our audit and the data collected. Next, in Section~\ref{Sec:Results} we present and analyse the results from the 101 audited papers, followed by a discussion in Section~\ref{Sec:Discussion} of the significance of our findings. Finally, we conclude with a summary of the key points arising from this analysis, various recommendations to the community along with consideration of its significance, weaknesses and areas for further research. 

\section{Background}\label{sec:background}
\noindent
In this section, we discuss the context of our SDP audit drawing from both software engineering and more widely in order to understand the backdrop to our audit and the motivation to undertake it.  

\subsection{Problems in Scientific Research}
\noindent
Back in 2005, in a highly controversial paper, Ioannidis \cite{Ioannidis2005} suggested that most scientific results were wrong. Subsequent studies have lent some support to this viewpoint, e.g., \cite{Earp2015,Nuijten2016,Heroux2023}. Diong et al.~\cite{Diong2018} conducted an audit of physiology and pharmacology papers comparing two time periods (2007-2010 and 2012-2015) and found extensive statistical errors and poor reporting practices with little evidence of improvement.  More recently (2023), from an analysis of papers published in psychology, Brewin~\cite{Brewin2023} reported that they were ``marred by multiple errors and inaccuracies and often fail to reflect the changing nature of the knowledge base''.    

Closer to our focus of machine learning applied to SDP, Kapoor and Narayanan~\cite{Kapoor2023} report there are many methodological pitfalls, including data leakage, essentially where the separation between training data and unseen validation data is violated.  The consequence is over-fitting leading to frequent over-optimism and over-claiming by researchers.  Once these problems are accounted for, they suggest that modern, complex methods frequently fail to out-perform simple benchmarks such as logistic regression.  Also addressing machine learning experiments---though focused on the domain of medical imaging---is the critique (rather than formal audit) by Varoquaux and Cheplygina~\cite{Varoquaux2022} who also note widespread methodological problems and sources of bias, in part arising from the desire to publish and the quest for novelty.   

Fazekas and Kov{\'a}cs~\cite{Fazekas2024} examinined results from machine learning experiments in an audit that looked at the consistency of reported classification performance metrics in medical image processing.  Typically, they found researchers report multiple metrics, many of which have structural relationships thus enabling consistency checks.  Notably, they identified inconsistencies in $\sim 30\% $ of papers from an audit of 100 highly-cited scientific papers.  

\subsection{Problems in Software Engineering}
Within software engineering, we can go back to a seminal paper in 2002 by Kitchenham et al.~\cite{Kitchenham2002} who undertook, to the best of our knowledge, the first critical review or audit of the empirical software engineering literature with a goal to assess statistical practice and make recommendations.  They concluded, that ``software researchers often make statistical mistakes''; however, the analysis was based on only eight, ``non-randomly sampled'' papers.  Nevertheless, this is a highly influential paper in addressing methodological quality in software engineering research.

Specifically, within SDP, pioneering work was undertaken by Bowes et al.~\cite{Bowes2014} who developed a checking tool DConfusion to assess the consistency of a confusion matrix which is a $2 \times 2$ matrix of true-positive, false-positive, false-negative and true-negative counts.  This forms the basis for the majority of classification metrics and thus the basis for consistency checking\footnote{The work by Bowes et al.~\cite{Bowes2014} underpins many of the ideas of the more recent audit by \cite{Fazekas2024}.}.  They illustrated the utility of DConfusion on a set of example papers and revealed consistency problems that were subsequently confirmed by the authors as indeed arising from a typographical error. 

These ideas were then deployed in a previous audit\footnote{Note that our audit extends the previous findings with an entirely new, larger sample, broadened scope and also examines the question of study reproducibility} by ourselves examining 49 papers drawn from a systematic review of unsupervised learning algorithms applied to SDP~\cite{Shepperd2019}. Disturbing findings included that $\sim45\%$~(22/49) papers contained demonstrable errors.  In addition, incomplete reporting on many occasions made it difficult to determine whether an error was present or not.

So to summarise, problems in scientific research papers abound, as has been revealed by multiple studies and audits.  Preliminary evidence suggests that the use of machine learning techniques for software defect prediction are not immune; consequently, we decided to undertake a larger and more extensive audit of recent (2019-2023) experiments.

\subsection{Reproducibility and Replicability}

At this juncture, we need to be clear that reproducibility is distinct from replicability, since assessing study reproducibility is one of the motivations for this audit. RQ3 explores how well SDP studies report on their methods and results in order to support reproducibility.

\begin{description}
\item[Reproducibility:] is the ability to recreate the results of a study by using the same data and following the exact same procedures as the original experiment. In terms of the ACM Badges definition \cite{ACM2020}, we mean both repeatable (by the same team) \textit{and} reproducible (by another team).  
\item[Replicability] is the ability to obtain consistent results across different studies, potentially by different researchers and under potentially different conditions. This may involve using different samples, locations or even slight variations in methods.
\end{description}

Reproducibility is therefore a necessary precursor to replicability \cite{Madeyski2017}. It is an important and often overlooked concept in empirical research that serves at least two purposes. Firstly, it promotes a high level of confidence in the results, that they are demonstrably correct.  Secondly, reproducibility facilitates future replication.  Specifically for software engineering, are assessment proposals from Gonz{\'a}lez-Barahona and Robles \cite{Gonz2012,Gonz2023} and also Liu et al.~\cite{Liu2021}.  We largely follow the original approach of \cite{Gonz2012} for reasons of simplicity and to not ``reinvent the wheel'' (see Section~\ref{SubSec:DataColl} for a more detailed description of our approach).

Related is a move towards Open Science or Open Research.  Clearly, undertaking such activities as sharing data, analysis methods and other research artefacts is likely to underpin reproducibility and replicability. Initial calls came from the wider scientific research community, e.g., \cite{Munafo2017} and were in part triggered by the widespread inability to replicate results ind disciplines such as psychology \cite{Open2015}. This led computer scientists and software engineering researchers to explore whether similar problems existed in our research fields.  An investigation into Open Science and the lack of shared artefacts in software engineering by Heum{\"u}ller et al.~\cite{Heumuller2020} of 789 ICSE research track papers between 2007 and 2017 showed a positive trend towards artefact availability, but even by 2017, only 58.5\% of studies made their research artefacts available.

To summarise, there is a growing sense within the software engineering community that Open Science and reproducibility of a study are important but not complete consensus as to what this entails and, clearly, still some way to go in achieving this.

\section{Method}\label{Sec:Method}

Our goal is to evaluate the quality of published computational experiments in the domain of software defect prediction. Since we want to consider recent and current practice, we focused on the past five years, i.e., 2019-2023.  

To achieve this, we undertook an audit, which Ralph and Baltes~\cite{Ralph2022} refer to as a critical review\footnote{We prefer the term audit, as it is more widely used beyond software engineering, e.g., in the context of statistical methods.}.  This is an evaluative process that assesses the quality, reliability and validity of existing primary studies in a specific field. It entails a detailed examination of the research design, methods, data analysis and findings of individual studies with the goal of identifying strengths, weaknesses, biases and gaps. So the purpose is to provide a comprehensive understanding of the state of research on a particular topic and to offer constructive feedback for improvement. 
 In software engineering, they have been deployed to investigate methodological topics. For example, Baltes and Ralph's critical review/audit \cite{Baltes2022} evaluates how software engineering researchers use and report on the representativeness of the sample of the target population.  
 
\begin{figure}[htp]
\begin{center}
\includegraphics[width=\columnwidth]{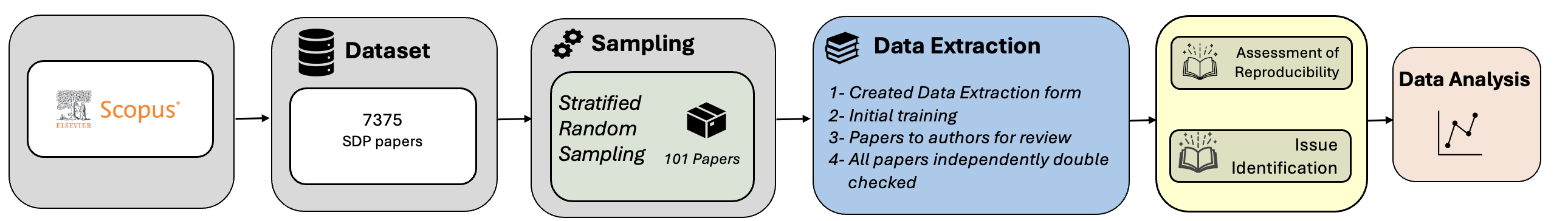}
\caption{Overall audit process}
\label{fig:OverallMethod}
\end{center}
\end{figure}

 Note that we are conducting a systematic technical review against established methodological standards rather than an audit against a particular pre-registered or published standard.  Note also that this is distinct from a traditional systematic review or meta-analysis, where the emphasis on evidence generated by primary studies; the goal in that case would be to synthesise all relevant studies on a specific research question to provide a comprehensive summary of the evidence. Low quality studies are typically excluded (or at least down-weighted). This contrasts with an audit, where we explore the quality of the studies, not the evidence. We summarise the process diagrammatically in Fig.~\ref{fig:OverallMethod}.  

Specifically, we use the audit to answer the following questions:
\begin{enumerate}
    \item Using bibliometric data, how can we characterise our sample of SDP studies?  
   \item What experimental design approaches are used in SDP research?  
   \item How reproducible are SDP studies?  
   \item What kinds of quality issues are found in SDP studies?  
\end{enumerate}

\subsection{Search}\label{SubSec:Search}

We chose Elsevier's Scopus database \cite{Baas2020scopus} for two reasons.  First, like Web of Science (WoS), it is subject to various quality requirements such as ensuring that all papers are subject to at least some minimal peer review process and has active policing of fake journals and so-called predatory publishers. Second, it has a broader scope than WoS and, in particular, better coverage of conferences.  As a result, our population excludes the so-called grey literature \cite{Kamei2021grey} and covers fewer venues than compared to a database such as Google Scholar. This was intentional, since our focus is on mature, refereed studies and what is hopefully, state-of-the-art practice.

For the actual search, we used the following, covering all fields, but restricted to: (1) articles and conference items, (ii) computer science, (iii) English items and (iv) years 2019-2023.

\begin{verbatim}
    ( "software defect" OR bug OR "software fault" ) 
    AND prediction 
    AND "machine learning"
\end{verbatim}

\noindent
This retrieved 7,375 documents (09/04/2024). Given the large numbers, we undertook a stratified random sample to select 20 papers from each year to achieve a target of 100 papers\footnote{Our total sample size was 101 due to the additional inclusion of an initial training paper.}.  Our rationale was that we were interested in changes over time and in particular whether there were any improvements. To achieve this, we randomised papers within year and searched until a target of 20 relevant papers satisfying all inclusion criteria was achieved. These were:

\begin{enumerate}
    \item a focus on predicting software defects
    \item presenting a new experiment with results
    \item based on real-world data as opposed to simulated data or student experiments
    \item availability of the full content (14 articles excluded)
    \item a minimum length of 3 pages 
\end{enumerate}

A common concern is whether 100 papers can ‘represent’ all SDP studies (2019-2023). However, our sample was probabilistically drawn from a fully enumerated sampling frame, ensuring that each article had an equal probability of selection. This allowed us to draw unbiased estimates of population-level properties with calculable margins of error.  

To estimate the proportion of the total number of relevant articles sampled from, we made the following calculations given in Table~\ref{Tab:SampleCalcs}.  Not all the 7,375 articles retrieved by the search were relevant.  In finding the required number (101) for the audit sample, we discarded 354 from which we could extrapolate to estimate the total number of relevant articles.  From this, we suggest that between 2019 and 2023 an \textit{estimated} 1,585 primary studies on experiments into SDP have been published and indexed by SCOPUS. Since SCOPUS does not cover all software engineering conferences, the true figure is likely to be greater than this. Our audit sample reviewed 6.4\% of these articles. 

\begin{table}
\centering
\caption{Sample Calculations}
\label{Tab:SampleCalcs}
\begin{tabular}{lr}
\hline
Articles                          & Count or proportion  \\
\hline
Total from SCOPUS                 & 7,375                \\
Articles manually searched        & 469                  \\
Irrelevant articles rejected      & 354                  \\
Unavailable articles rejected     & 14                   \\
Articles in final sample          & 101            \\  
Hit proportion (101/469)          & 0.215                 \\
Estimated total relevant articles ($7375 \times 0.215$) & 1,585       \\
Sample proportion (101/1585)       & 6.37\% \\   
\hline
\end{tabular}
\end{table}

\subsection{Outcome variables}\label{subsec:outcomevars}

Our audit addresses two outcome variables: reproducibility and issues (which are split into experimental design plus implementation issues and reporting issues).

\subsubsection{Reproducibility metrics}

To assess the reproducibility of each paper, we adapted the Instrument proposed by Gonz{\'a}lez-Barahona and Robles~\cite{Gonz2012} and revisited in 2023~\cite{Gonz2023}. Our adaptation was to simplify that aspect of the instrument related to the raw data collection due to all experiments in our audit being based on secondary data.  Gonz{\'a}lez-Barahona and Robles' approach has the merit of being quite generic and simple to apply since the questions are all either yes or no. We took the view that all questions should contribute equally and then normalised the total score to give a range between zero and one.

There are 27 basic indicators scored (0 or 1) e.g., indicating how easily the element, e.g., a data set, can be identified for replication (see the Appendix for full details). These fall into five categories: 
\begin{enumerate}
\item Identification: is the element actually specified or identified?
\item Description: The detail level of the published information about the element.
\item Availability: Indicating how easily the element can be obtained.
\item Persistence: Indicating the likelihood of the element being available in the future.
\item Flexibility: How easily the element can be adapted to new environments.
\end{enumerate}

Of course, there is some subjectivity in scoring, especially for borderline cases. We believe the use of two independent reviewers and the fact that are 27 items ameliorates this problem, (revisited in Section~\ref{Sec:Threats} Threats to Validity).  These categories are then applied to five aspects of a study, namely:
\begin{enumerate}
    \item Raw data
    \item Extraction methodology/tools
    \item Processed dataset				
    \item Analysis methodology/tools
    \item Results dataset
\end{enumerate}

\noindent
The remaining two indicators relate to the specification of relevant hyperparameters. This results in an overall Reproducibility Score out of 27 then normalised to 0-1.

\subsubsection{Study Issues Assessed}\label{subsubsec:issues}

This section defines and explains the nine issues we checked for in each article. These fall into two groups. Issues 1–5 concern the design and conduct of the experiment, while Issues 6–9 concern reporting. We selected issues that are widely relevant, detectable, non-trivial and, as far as possible, uncontentious. For this reason we avoided more debatable topics such as the use of p-values in NHST \cite{Benavoli2017}. We also aimed to be sensitive to context, since different studies have different goals that may call for different design choices. At the same time we recognise that, although a range of practices persists in SDP, the broader statistics and machine learning communities have shifted toward more robust and transparent approaches. In short, this is an audit rather than a critique, and our aim is to be flexible and only flag issues that threaten the validity of an SDP experiment or limit its scientific value.

\begin{enumerate}
\item Lack of Out of Sample (OOS) validation: Distinct from model-fitting (using all available data), SDP studies seek to evaluate the predictive performance of the learner where it deployed upon unseen data. This is accomplished by simulating the process of using a trained learner to predict the target variable of a new and unseen instance. OOS validation strategies commonly used in machine learning experiments include:
\begin{enumerate}
    \item  Holdout / train–test split: The dataset is randomly split into separate training and testing sets and the performance of the held out test set reported.   Typically this is in a fixed ratio (e.g., 70\% training, 30\% testing) \cite{Provost2001}  Clearly such a procedure is vulnerable to the chance allocation of individual instances.
    \item k‑fold Cross‑Validation (CV): Since the 1990s the standard approach for OOS validation has been cross-validation \cite{Kohavi1995}. This works by randomly allocating instances to $k$ approximately equal-sized folds.  One fold is then used as the validation or `unseen' set and the remaining $k-1$ folds are used for training. This procedure is repeated so that after $k$ iterations every instance has been entered once into the validation fold. Kohavi recommends $k=10$, although there appears some flexibility in practice.  Some researchers advocate the use of $j \times k$ cross validation where the whole procedure is repeated $j$ times to reduce errors in the estimate of the mean prediction metrics.
    \item Leave-One-Out Cross-Validation (LOOCV): Each instance in the dataset is used once as a test set while the remaining instances form the training set \cite{WongYeh2019}.  This approach is deterministic but can be computationally very demanding.
    \item Bootstrap Resampling: This involves repeatedly sampling with replacement of the dataset to create training and testing sets. It provides robust estimates of model performance \cite{Kohavi1995}.
    \item Cross-Project Validation: Training is performed on one project and testing is done on a different project in order to evaluate generalisability across projects \cite{Menzies2012}.
    \item Time-Based Validation: Used in scenarios like Just-in-Time (JIT) defect prediction, where the training and testing sets are divided based on temporal order to simulate real-world deployment \cite{Stapor2021}.
\end{enumerate}
Clearly some strategies are more appropriate/realistic for SDP such as cross-project and time-based, however, for the purposes of our audit we merely determine whether OOS validation has been undertaken.
\item Using problematic metrics: this is particularly relevant for classifiers where a number of researchers in multiple fields have convincingly demonstrated that some widely used performance metrics such as F1 and accuracy are biased and unreliable for two-class classification problems (see, for instance \cite{Powers2011,Chicco2020,Yao2021,Lavazza2022}). Specifically, Accuracy is inadequate under skew as majority class dominance can yield deceptively high scores even for trivial predictors.  F1 was proposed for information retrieval problems \cite{vanR79} which are generally 1-class, e.g., the number of relevant pages correctly retrieved whilst the number of irrelevant pages not retrieved is both unknowable and uninteresting.  For 2-class problems it is vulnerable to bias and is difficult to interpret.  Correlation metrics, e.g., MCC uses all four cells (TP, TN, FP, and FN) and are typically less biased than Accuracy or F1 on imbalanced data. However, MCC is \emph{not completely unbiased} and can degrade under \emph{extreme} skew \cite{zhu2020mcc}.  
To continue to use such metrics is problematic.  We did however, check the specific use and context of problematic metrics for each paper so, for instance, where a metric was reported but flagged as potentially misleading, this was \textit{not} recorded as an issue.
\item Benchmarks and better than random: another problem with predictive performance metrics is the need for some benchmark to establish that the learner is doing better than guessing. Metrics such as Area Under the Curve (AUC) and Matthew's correlation coefficient (MCC) are chance-anchored \cite{Chicco2021}.  Youden's $J$ (Bookmaker odds / Informedness) provides another chance-anchored view at a chosen threshold and is a principled alternative when an operating point must be fixed, e.g., when using logistic regression.  Failure to include such a benchmark is again problematic. For the record, Yao and Shepperd~\cite{Yao2021} showed that 16 out of the 33 reviewed studies contained at least one result worse than random, that is, it was a perverse predictor. 
\item Multiple statistical tests without correction: computational experiments such as SDP studies typically generate many results $(\#Learners \times \#Datasets \times \#Metrics $ and these are often evaluated using statistical inference tests. Again, it is well known that without making some kind of correction, the false positive rate will be greatly inflated \cite{Midway2020}.  In the spirit of only focusing on clearly problematic issues/methods we accept \textit{any} method to correct alpha as being acceptable although obviously a more modern approach is preferable e.g., Benjamini-Hochberg \cite{Benjamini1995}.
\item Addressing spread as well location: typically results are given as a single statistic which only summarises the central tendency of the mean predictive performance. This reduction of many results to a single summary statistic has some ramifications, such as the extent to which results may vary due to the stochastic nature of both algorithms and validation procedures being unknown.  Boxplots of the performance metrics are an effective, and quite widely used graphical method to indicate dispersion.
\item No link to data: the data used is not fully indicated e.g., version or location. 
\item No link to code: the relevant code is not shared. 
\item Low reproducibility: the level of reporting for reproducibility is unacceptably low, which we define as a score of less than 50\% for the Gonz{\'a}lez-Barahona and Robles~\cite{Gonz2012} instrument meaning that 14+ out of 27 questions are answered `No'.
\item Article behind a `paywall': Given the widespread opportunities to provide a final postprint version of the paper in a publicly accessible, not-for-profit archive, not doing so seems both perverse and harmful to the progress of research.
\end{enumerate}

\subsection{Data collection and analysis}\label{SubSec:DataColl}

For the audit, all authors independently reviewed one paper which we then collectively discussed and resolved disagreements. The guidance notes and definitions were then refined. Each author was then allocated two blocks of 14-15 papers, meaning that all papers were independently read and reviewed by two authors and again disagreements resolved. This process enabled us to extract the following data (for full details refer to the Appendix).

\noindent
This information covered the following areas:
\begin{enumerate}
\item Bibliographical details: publication venue, year, page count and citation information of the research paper being analysed.
\item Open Access: indicates the type of access e.g., Gold, Green, etc.
\item Document Type: specifies if the publication is a journal or conference article.
\item Experimental design information including number of datasets, learners and choice of performance metrics and out of sample validation methods. 
\item Choices of benchmarks e.g., comparison with a random predictor.
\item Statistical methods including how the results are summarised, type of inference and statistical tests.
\item Reporting details and reproducibility score.
\end{enumerate}

Note that the data analysis was performed using R embedded in an RMarkdown Notebook which is shared on zenodo along with our data.

\section{Results}
\label{Sec:Results}

In this Section we present the results of our audit in order of the four research questions.

\subsection{RQ1: Using bibliometric data how can we characterise our sample of SDP studies?}

\begin{figure}[htp]
\begin{center}
\includegraphics[width=\columnwidth]{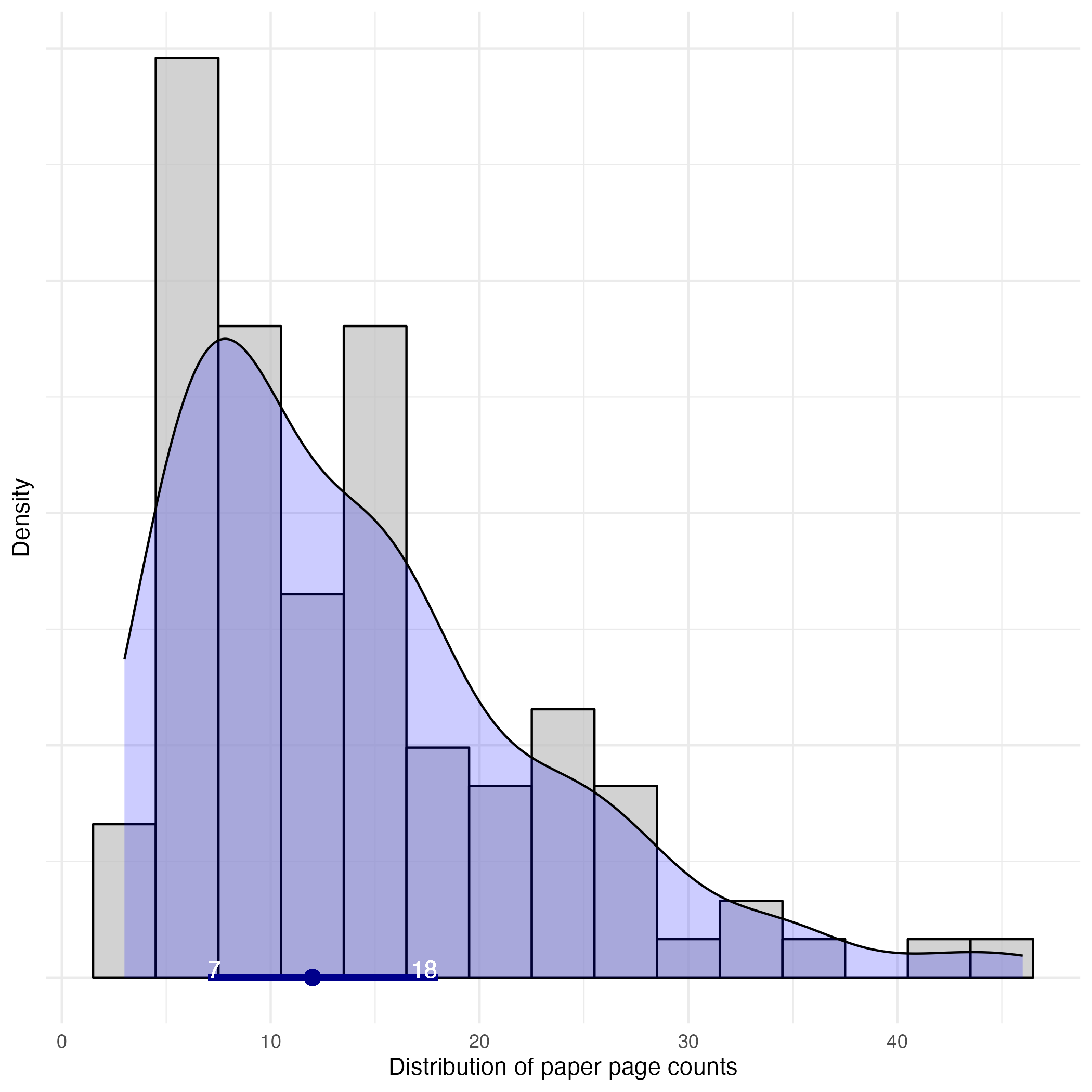}
\caption{Distribution of paper length by page count}
\label{fig:PageHist}
\end{center}
\raggedright{{\footnotesize The horizontal blue line along the x-axis shows the interquartile range, with Q1 and Q3 values given in white and the median represented as a blue circle. Thus, in this distribution, 50\% of the papers are between 7 and 18 pages in length.}}
\end{figure}

There are 40 conference and 61 journal papers in our sample from 2019-2023. A conspicuous feature of our sample is the diversity of sources, i.e., distinct journals and conferences. The 101 papers cover 74 unique sources or venues. This underscores the scale of research activity, the distributed nature of publications and the challenges of maintaining a comprehensive and up to date understanding of the field.  
Next, we examine paper length which ranges from 3 to 46 pages. The minimum is an result of our decision to exclude any paper shorter than 3 pages from the audit, on the grounds of it not having sufficient space to fulfil the norms of scientific reporting. The median length is 12 pages, reflecting the page restrictions typically imposed by conference publication (See Fig.~\ref{fig:PageHist}).

Then we looked at whether papers were open access or not.  Aside from moral or ethical considerations, there is also growing evidence that this has a marked impact on the diversity and volume of citations from other researchers \cite{Huang2024}. From our analysis, we see that 50/101 i.e., just under 50\% of papers are behind paywalls and therefore open access (see Table~\ref{tab:AccessCts}).  
 
\begin{table}
\centering
\caption{Counts of Paper Access Types}
\label{tab:AccessCts}
\begin{tabular}{lrr}
Paper Access Type & \multicolumn{1}{l}{Count} & \multicolumn{1}{l}{\%} \\
\hline
Gold/diamond& 29     & 28.7  \\
Green       & 23     & 22.8  \\
Green (Not available)  & 41     & 40.6 \\
No          & 8      & 7.9 \\
\hline
Total       & 101 & \\                  
\end{tabular}
\end{table} 

All the mainstream software engineering publishers e.g., Elsevier, IEEE and Springer allow the author to post a postprint, meaning Green Access. What is frustrating is that for most Green published papers (41/64), authors have chosen \textit{not} to make a postprint available (for example on their own institution website or a public, not-for-profit archive such as arXiv or zenodo).  

Gold or Diamond Access is when either the author’s institution pays the publisher either as optional charges (e.g., Elsevier) or sometimes as a compulsory article processing fee (e.g., PLOS ONE, Frontiers, etc.); this allows the article to be found on the publisher’s website and freely downloaded by anyone.  For more detail and explanation see~\cite{Harnad2004,Meagher2021}.

\begin{figure}[htp]
\begin{center}
\includegraphics[width=\columnwidth]{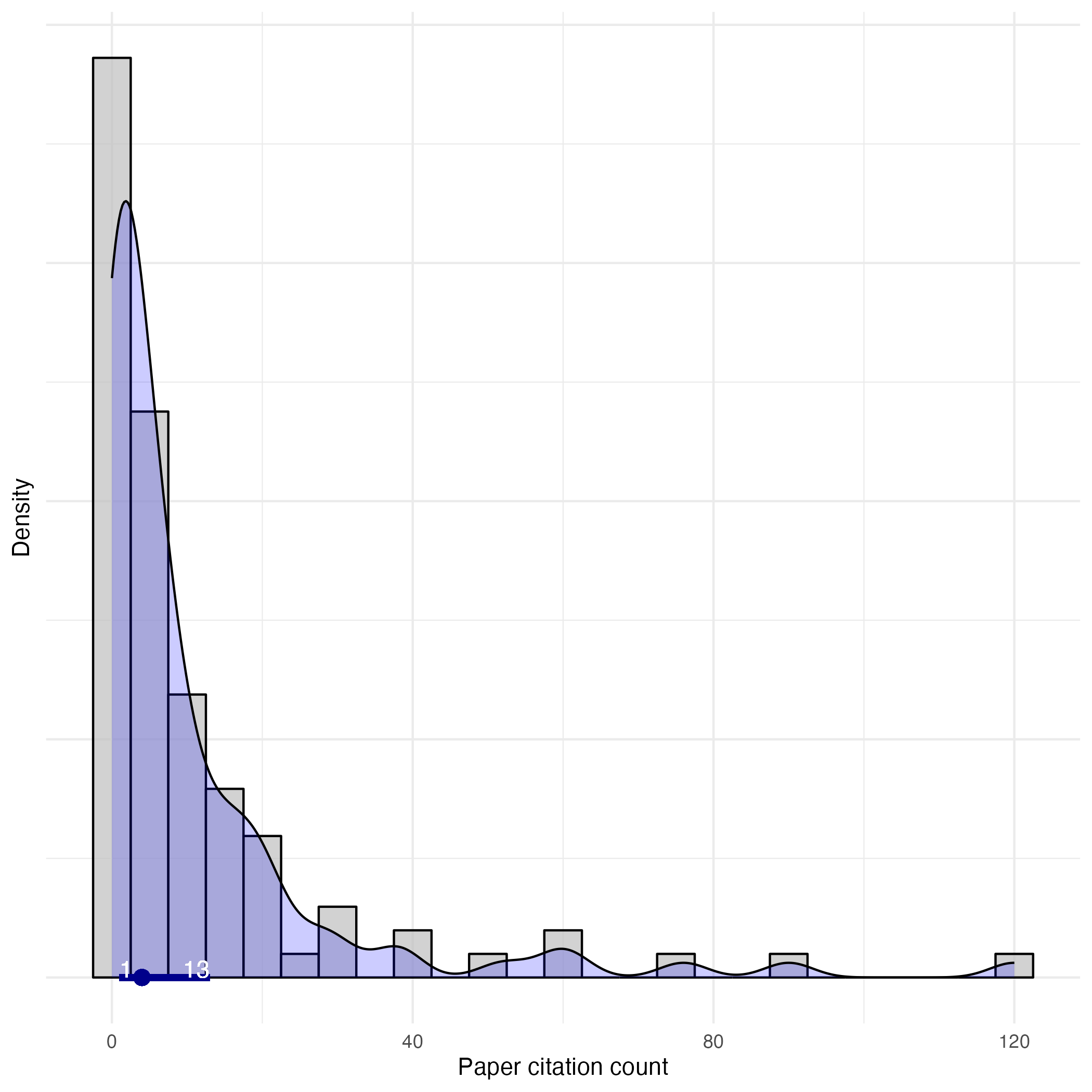}
\caption{Distribution of paper citation counts (unnormalised)}
\label{fig:CiteHist}
\end{center}
\end{figure}

We also investigated paper citation information (see Fig.~\ref{fig:CiteHist}). This reveals a strong positive skew (not least because negative values are impossible for count data) ranging from 0 to 120, with a median of 4. The IQR of 1-13 is highlighted on the x-axis of the histogram as a dark blue bar. The high variance was somewhat surprising (to us) nor did it particularly fit our expectations of what might be thought of as seminal papers. We investigated self-citations which were generally too low (0-10 with a median of zero) to explain the variability.  

\begin{figure}[htp]
\begin{center}
\includegraphics[width=\columnwidth]{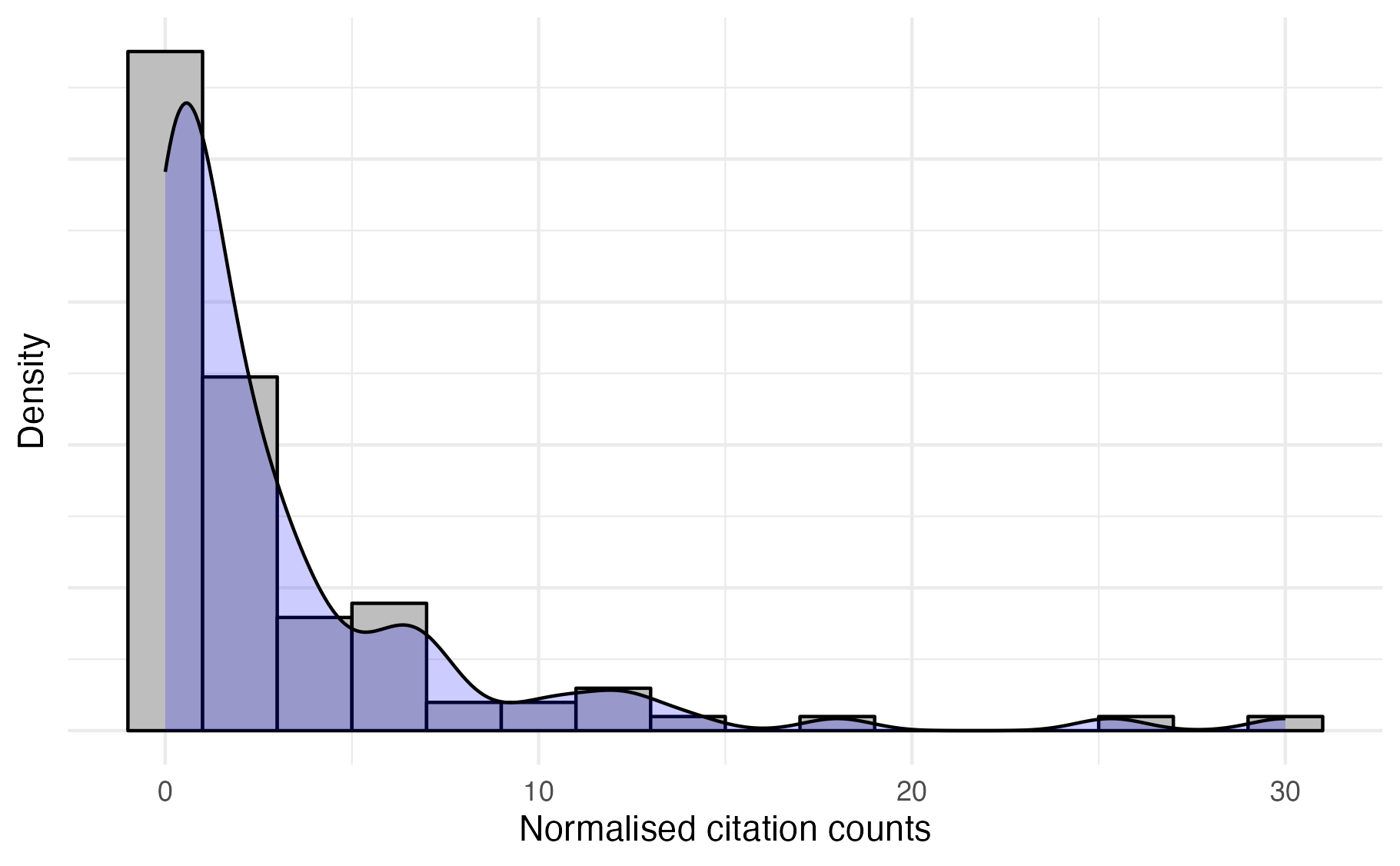}
\caption{Distribution of paper citation counts normalised by year}
\label{fig:CitesNormalisedHist}
\end{center}
\end{figure}

We next normalised the citations by number of years, since publication (see Fig.~\ref{fig:CitesNormalisedHist}); there remains high variability ranging from 0 to 30. The median for all years is only 1.2 citations per year and indeed 25/101 papers have no citations.  %

Next, we ask the question: does publication type influence citations?  Given the highly skewed distributions, we compare medians and the 95\% confidence intervals (see Table~\ref{tab:CitesPerYrByDocType}). There is some evidence of an effect since the median citation rate per year for a journal paper is 2.2 compared with 0.78 for a conference paper; however, the 95\% confidence intervals overlap in part due to the high variance of paper citation counts, even when normalised.

\begin{table}
\centering
\caption{Median Citations per Year by Paper Type}
\label{tab:CitesPerYrByDocType}
\begin{tabular}{lrrr}
Paper Type & Median & Lower Bound & Upper Bound \\
\hline
Conference & 0.775 & 0.60 & 1.33 \\
Journal    & 2.200 & 1.00 & 3.40  \\
\hline              
\end{tabular}
\end{table} 

We also observed two papers containing multiple examples of tortured phrases i.e., ``unexpected, weird phrases in lieu of established ones'' \cite{Cabanac2021} such as, in our audit, ``novel insects'' instead of ``new bugs'' and ``arranging of deformities'' instead of analysis of defects. Such phrases are usually indicative of problematic papers arising from the desire to evade plagiarism checkers by translating text from English to a second arbitrary language, sometimes multiple times and then finally translating back to English. Typically, these tactics are deployed by ``paper mills'' or other ``bad actors'' to create fake scientific papers \cite{Richardson2025}. Apart from the concern that the scientific body of work is being contaminated by these meaningless papers, it suggests a lack of oversight by the community and indeed both papers have been cited by others. 

Since the tortured phrases occur from the very start of each paper, it does cause us to wonder how carefully they have been read.  For this reason we unsuccessfully approached the authors, editors and publishers recommending they be retracted. The papers concerned are (i) ``Machine Learning-Based Defect Prediction for Software Efficiency'' published in 2023 in \textit{The International Journal of Intelligent Systems and Applications in Engineering} and (ii) ``Software Defect Prediction Framework Using Hybrid Software Metric'' published in 2022 in \textit{The International Journal on Informatics Visualization}. We also informed Scopus who responded that they no longer index \textit{The Intl.\ J.\ of Intelligent Systems and Applications in Engineering} and are investigating (2/10/2024) the International Journal on Informatics Visualization.

 It is hard to estimate the overall prevalence of papers containing tortured phrases. Apart from the two papers we located on our sample of 101 papers, by using the Problematic Paper Screener curated by Cabanac et al.~\cite{Cabanac2022}, we located a further 9 contained in our initial Scopus search of 7,375 papers. This is likely to be an underestimate because it is hard to anticipate a tortured phrase until it is encountered and Cabanac's background is the scientific literature more generally.  However, using the two positive observations from a sample of 101, the exact Clopper-Pearson confidence interval, which is generally more reliable for small sample sizes or proportions close to 0 or 1 gives a 95\% confidence interval of [0.002, 0.070]. In other words, a prevalence of between 0.2\% and 7\%.

So, to summarise, there seem disappointing levels of open access, much variability in citation patterns and in paper length and weak evidence that journal articles tend to be more highly cited. Worryingly, even in a carefully curated research paper collection such as Scopus, we see clear evidence of domain-incoherent terminology and suspect papers.

\subsection{RQ2: What experimental design approaches are used in SDP research?}

First, we categorised papers by whether the prediction system under experimental investigation was a classifier --- in practice a dichotomous classifier --- or a regression system predicting fault count or severity.  As it transpired, almost all studies focused on classifiers, although sometimes as part of a larger objective such as the effort-aware approaches. Only 4 out of 101 studies directly used regression systems to predict continuous valued outputs.  It should be noted that regression systems naturally lead to different types of prediction accuracy metric; however, for the purposes of this audit, we do not differentiate between them and classifiers and use the more generic term ``learner''.

We observed considerable variability in experimental design.  Table~\ref{tab:ExptDesCts} shows summary statistics for the number of datasets (this includes new releases or versions of a system), distinct learning algorithms and performance evaluation metrics.  If there is such a thing as a typical study, then it evaluated 5 learners over 10 datasets or system versions and used 3 performance metrics to make comparisons. Typically, the results were presented in tables, so for our hypothetical study, this would result in $5 \times 10 \times 3 = 150$ cells or results from which preference relations can be established.

\begin{table}
\centering
\caption{Experimental Design Variables}
\label{tab:ExptDesCts}
\begin{tabular}{lrrrrrr}
Count & Min & Q1 & Median & Mean & Q3 & Max \\
\hline
Datasets & 1 & 6 & 10 & 22 & 19 & 365 \\
Learners & 1 & 3 & 5 & 6 & 7 & 34 \\
Metrics & 1 & 2 & 3 & 4 & 5 & 9 \\
\hline            
\end{tabular}
\end{table} 

Almost all studies used secondary datasets. The number ranged from 1 to a remarkable 365 with a median of 10. The Promise repository datasets dominated, with it seeming to serve a similar role to the UCI datasets used in machine learning experiments more generally.  

Data pre-processing strategies also varied greatly.  One important area is how to address the challenge of imbalanced training data in that the positive cases (i.e., defective components) are almost always in the minority. Typical strategies include SMOTE \cite{Fernandez2018}. Overall, 42/97 studies explicitly used some imbalanced learning mechanism. We exclude regression studies since the notion of imbalance is not relevant to continuous target variables.  Given the almost universal use of imbalanced datasets, this proportion appears low. However, we decided not to treat it as a quality issue since it depends on the specific purpose of the SDP experiment and optimising the prediction is not always the goal.

Similarly, there is significant variation in the number of learning algorithms\footnote{In practice, almost all papers looked at classifiers; however, there are also four regression-based predictors aimed at continuous-valued outcomes so we use the more general term.} investigated. The median is 5 though the maximum was 34. There is a certain degree of subjectivity regarding whether some pre-processing, e.g., feature subset selection, constituted an additional algorithm or was essentially cleaning the data prior to the computational experiment. To determine this, we tried to follow the structure of a paper and, in particular, how the results were presented and interpreted.  

We also see some variability in the number and choice of prediction accuracy metrics e.g., accuracy and AUC. These form the outcome variables for the experiments.  There is a strong preference for employing multiple metrics, with a median of 3 and a maximum of 9.  Of course, a challenge is the situation when the multiple metrics are discordant \cite{Yao2021} and how the paper's results should be interpreted.  

Next, we consider the specific choices of accuracy performance metrics. These are given in Table~\ref{tab:MetricCts}. Note that the percentages are of 97 papers since the remaining four deal with regression systems.  We see that the F1 metric is still the most widely used metric despite considerable adverse criticism, e.g., \cite{Hand2018,Yao2021}.  Unsurprisingly, since they are constituent parts of F1, Precision and Recall are also widely used.  Area under the [ROC] Curve (AUC) \cite{Provost2001} is also used in half of the experiments we audited, but this metric conceptually differs from other metrics like F1, as it evaluates the performance of a classifier across a spectrum of decision thresholds, rather than at a single fixed threshold.  

Given the well known and widespread criticisms of Accuracy as a performance metric for imbalanced datasets, which is almost invariably the case in the domain of software defect prediction, it is surprising that more than a third (38/97) of experiments still use this metric. Finally, the uptake of the more recently advocated Matthews Correlation Coefficient (MCC) \cite{Baldi2000,Chicco2020} remains quite low at under 20\%.  

\begin{table}
\centering
\caption{Frequency of Accuracy Metric Usage}
\label{tab:MetricCts}
\begin{tabular}{lcrr}
\hline
Metric      & Chance & Uses & Percentage (n=97)  \\
& Anchored & & \\
\hline
F1          & No & 61    & 62.9 \\
AUC         & Yes & 53    & 54.6 \\
Recall      & No & 52    & 53.6 \\
Precision   & No & 43    & 44.3 \\
Accuracy    & No & 38    & 39.2 \\
MCC         & Yes & 17    & 16.8 \\
Specificity & No & 10    & 10.3 \\
Others & n.a.\ & 86 & n.a.\    \\
\hline
\end{tabular}
\end{table}

Of these metrics only MCC and AUC are chance-anchored, in that guessing will give predictive accuracy scores of 0 and 0.5 respectively (irrespective of the prevalence of the positive case). Consequently, it is possible to see how a classifier is performing with respect to a random strategy. In our audit, just under 60\% (58/97) of experiments use metrics that allow us to compare the results with guessing. With other metrics such as the popular F1-measure, it is not possible to know how a result compares with simply guessing \cite{Yao2021}. We return to this in the Discussion Section.

Next, we turn to the deployment of Out of Sample validation strategies. Table~\ref{tab:CVCts} summarises the range and frequency of approach. The NAs arise when essentially the style of algorithm or experiment means the test and training divide is fixed e.g., time order as in Just-in-time (JIT) prediction.  The Unclear category arises when the checkers are unable to determine whether an OOS-validation strategy has been used and indicates problems with how the experiment is reported.  

\begin{table}
\centering
\caption{ Out of Sample validation strategies deployed}
\label{tab:CVCts}
\begin{tabular}{lll}
\hline
 OOS validation Employed & Count & Percentage  \\
\hline
Yes          & 66    & 65.3  \\
No           & 24    & 23.8  \\
Unclear      & 7    &    6.9 \\
Not relevant (NA) & 4  & 4.0 \\
\hline
\end{tabular}
\end{table}

Where relevant, we also collected information on the number of folds and number of repetitions.  The modal fold count is 10 (which is quite widely advocated in machine learning studies) and the number of replications ranged from one (in 26 studies thus the most common) to 100 (in 6 studies).

Where there was some element of repetition, either through multiple folds or replication (e.g., from repeating a bootstrap) we noted how the results were presented. In almost all cases, a measure of location, usually a mean, but sometimes a median, was used.  Less typically, in 32/83 of relevant cases a measure of the dispersion of the results was provided, e.g., graphically using boxplots or a summary statistic such as standard deviation. We view this as necessary reporting since it is important to know how much predictive performance can vary as well as the most typical value.  

In terms of inferential mechanisms, $\sim 45\%$  of studies (46/101) made use of statistical tests based around the idea of comparing a p-value with a pre-determined alpha value or acceptance threshold. When the p-value is less than alpha it is declared as statistically significant. Of these 46 studies, only 60\% (28/46) made adjustments to alpha when conducting multiple tests \cite{Bender2001}.  There are philosophical and methodological ramifications, e.g., \cite{Colquhoun2014,Greenland2016}, but, suffice to say, not adjusting alpha in the context of multiple tests is problematic. This is because there are typically many inferential tests (the median is 24 and the maximum is 704). 

\subsection{RQ3: How reproducible are SDP studies?}

\begin{figure}[htp]
\begin{center}
\includegraphics[width=\columnwidth]{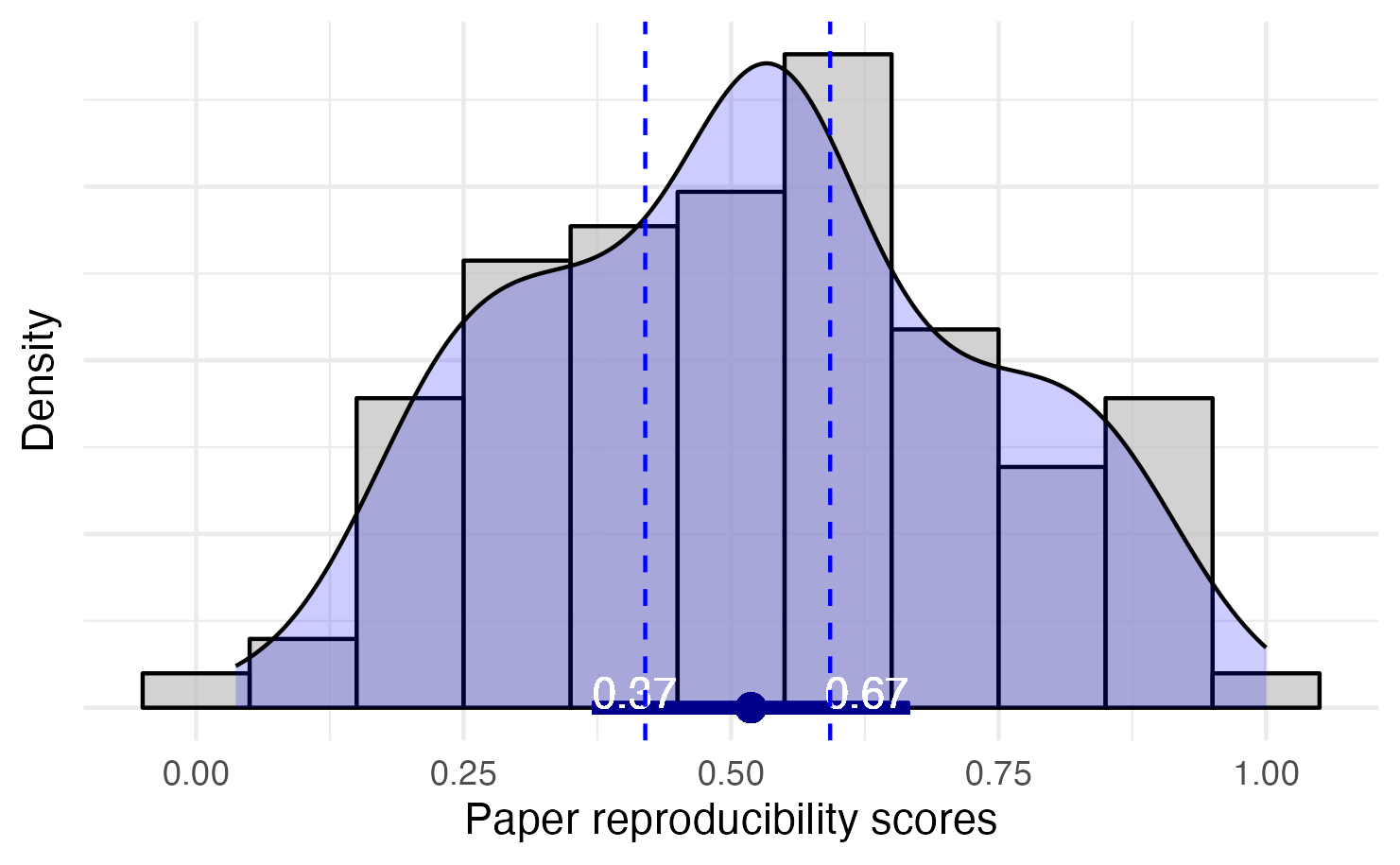}
\caption{Histogram and density plot of paper reproducibility scores}
\label{fig:ReproHist}
\end{center}
\raggedright{{\footnotesize Additionally to the IQR, the vertical blue dashed lines represent the tertile boundaries.}}
\end{figure}

As described in Section~\ref{Sec:Method}, we used a variant of the Gonz{\'a}lez-Barahona and Robles~\cite{Gonz2012} Reproducibility Instrument which generated a score from 0-27 which we then normalised between zero and one. We found scores taking almost the entire range of possible values from 0.04 to a perfect 1.0, with a median of 0.52.  Fig.~\ref{fig:ReproHist} shows the distribution of reproducibility scores.  

We used the odds-ratio derived from the contingency table (see Table~\ref{tab:reproTertiles}) to compare the lower and upper tertiles of conference and journal papers following a procedure recommended by Gelman and Park~\cite{Gelman2009}. We therefore have the odds of a journal paper falling into the top tertile for reproducibility, as opposed to the bottom as being 20/7 and compare this with the same odds for conference papers of 14/23. From this, we can construct the odds ratio which is 4.694 with 95\% CI [1.582, 13.923]. Although the interval is broad, it does not straddle unity and hence is supportive of journal papers being more reproducible than conference papers.

\begin{table}
    \centering
       \caption{Contingency table of Paper type by reproducibility score tertile}
      \label{tab:reproTertiles}
      \begin{tabular}{l|rrr}
      \hline
         & Reproducibility  & ~ & ~ \\ 
        \hline
        Paper type & Tertile 1 & Tertile 2 & Tertile 3 \\ \hline
        Conference & 20 & 13 & 7 \\ \hline
        Journal & 14 & 24 & 23 \\ 
        \hline
    \end{tabular}
\end{table}

A common and useful way to assist with reproducibility is to link to the data or code.  From the audit sample we found $67/101 = 66.3\%$ of studies contained a link or specific reference to the data used. Note that we excluded vague references to a general repository or where the version was unclear. Only $19/101 = 18.8\%$ of studies included links to the analysis code or scripts. Of the 67 papers that contained one or more links, in 24 cases (35.8\%) at least one link was broken.

\begin{figure}[htp]
\begin{center}
\includegraphics[width=\columnwidth]{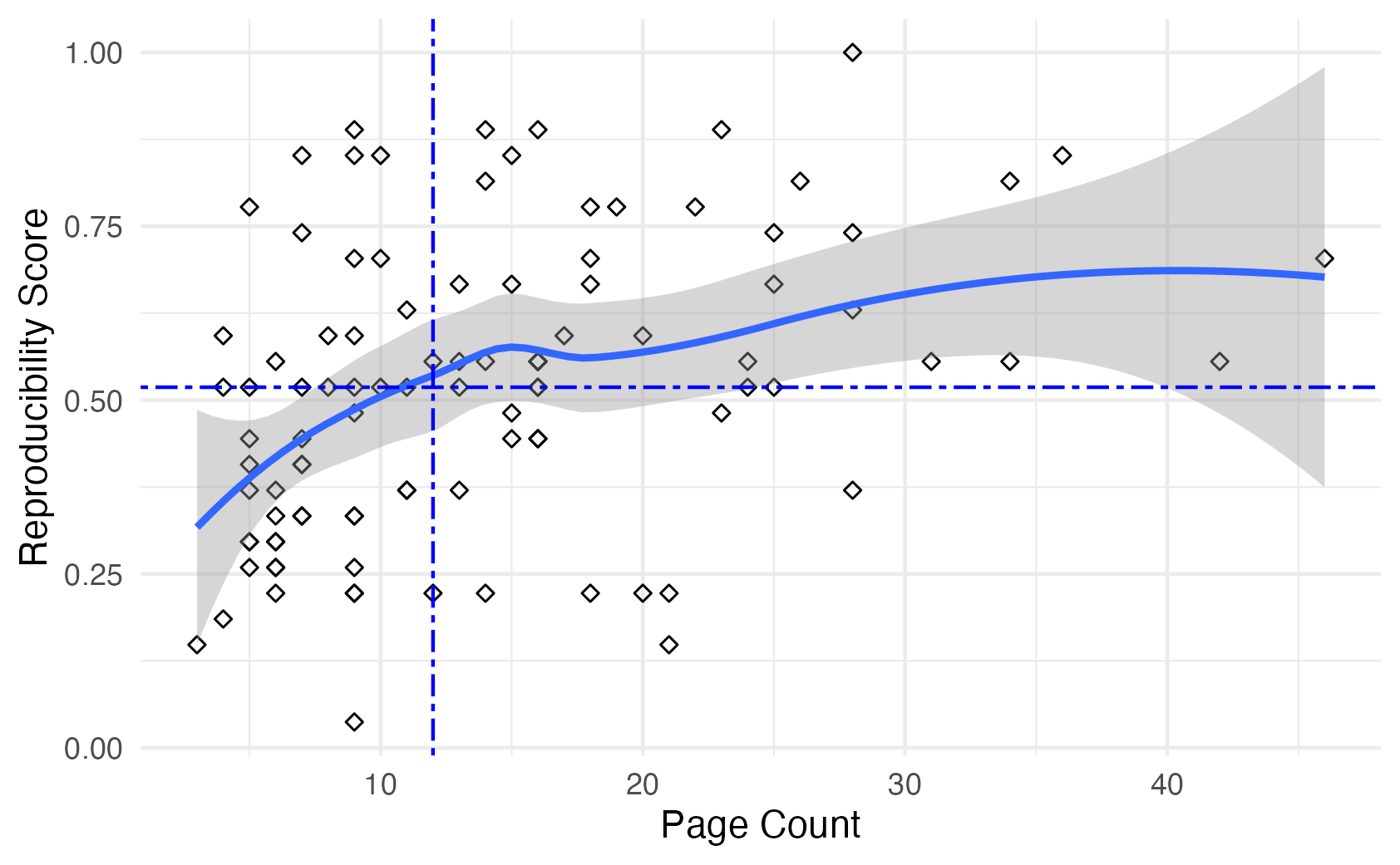}
\caption{Scatterplot of Reproducibility vs Paper Length}
\label{fig:PageReproScatter}
\end{center}
\raggedright{{\footnotesize The blue fit line is a loess smoother with associated 95\% confidence interval in grey. The blue dashed lines represent median values.}}
\end{figure}

Next, we explore some associations with, and \textit{possible} contributors to, reproducibility.  First, consider the relationship between paper length (in pages) and reproducibility. Here we might expect shorter papers to be less reproducible since there is less space to convey relevant information. As expected, there is a positive relationship (the Spearman correlation coefficient is 0.42) but there is a lot of scatter and an inflexion point around the median of 12 pages (see Fig.~\ref{fig:PageReproScatter}). However, there are clearly many other factors and inspection of the scatterplot suggests many shorter papers score highly on reproducibility and vice versa. Potential reasons include different journal and conferencing formatting styles. This means that a page does not contain a uniform number of words and author preferences in what information to communicate.

\begin{figure}[htp]
\begin{center}
\includegraphics[width=\columnwidth]{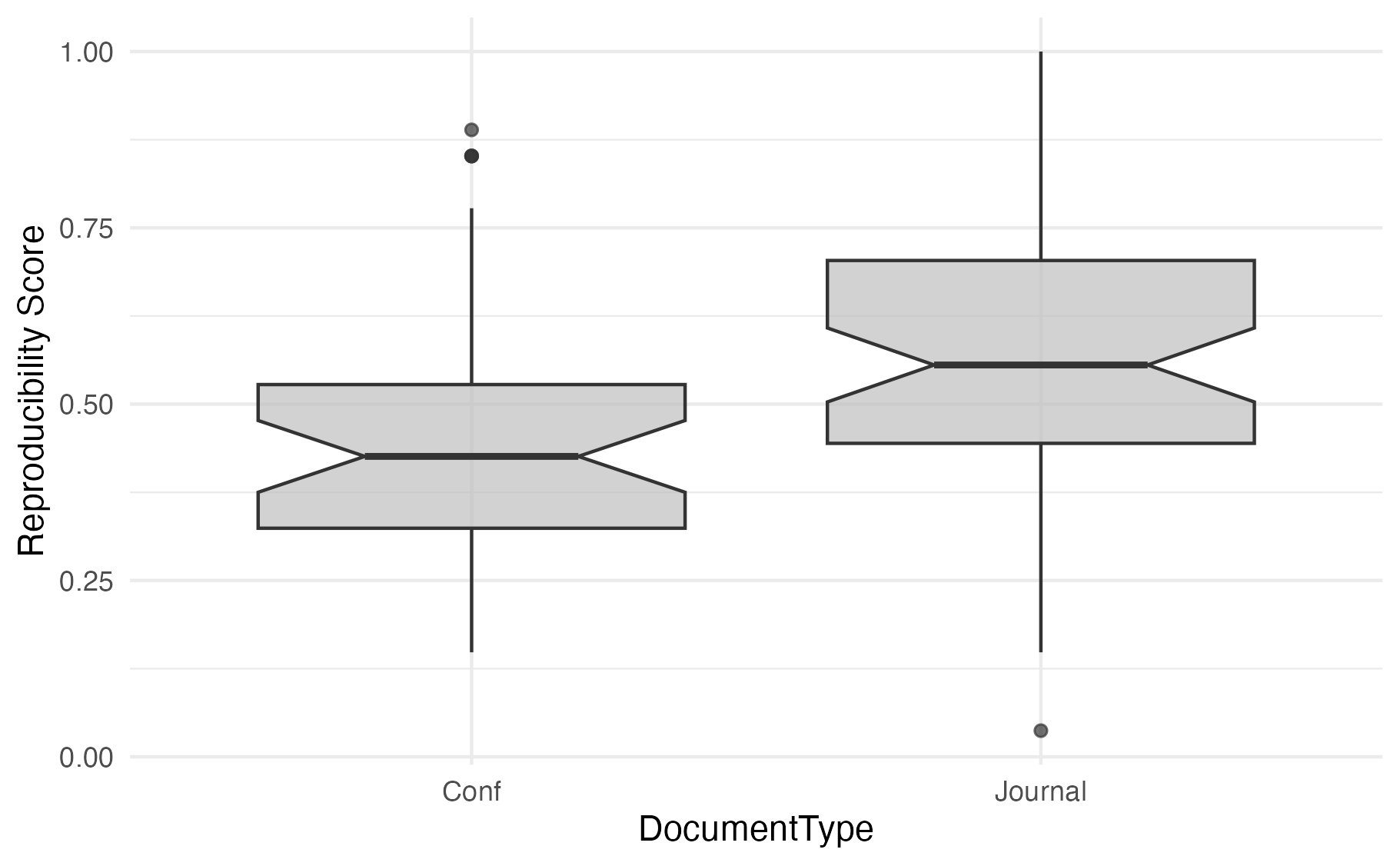}
\caption{Boxplots of Reproducibility vs Document Type}
\label{fig:DocTypeBoxplots}
\end{center}
\raggedright{{\footnotesize The notches show a parametric estimate of the median 95\% confidence intervals that assumes near normality.}}
\end{figure}

The side by side boxplots in Fig.~\ref{fig:DocTypeBoxplots} allow us to compare reproducibility between conference and journal papers. The medians are 0.426 and 0.556, respectively. In terms of effect size, this approximates to 3.5 additional positive answers out of 27 reproducibility questions. Using a robust method (5,000 bootstrap samples) to estimate the 95\% confidence intervals for the medians, we find they touch (Conference = [0.352, 0.519] and Journal = [0.519, 0.593]. The method used by ggplot2 in the boxplots makes more assumptions about normality and has tighter bounds. Either way, there is some evidence that journal papers may be more replicable than conference ones\footnote{This small distinction between methods of estimating confidence intervals is another reason we chose to avoid traditional significance tests since; in this instance, the choice of method determines whether the relationship is `significant' or not. We believe this to be a false dichotomy and prefer to focus on strength of evidence and estimating effect size, as opposed to declaring whether an association is true or not \cite{Greenland2016}.}.
We conjecture that the combination of more permissive page lengths and an iterative peer review process can lead to published experiments that are easier to reproduce. Of course, there are some conference papers that are clearly superior to the weaker journal papers; nevertheless, there is some pattern, which in terms of our conjectured causality, makes sense\footnote{Note that although our modelling approach is not directly causal since we explore association, nevertheless, awareness of potential causal mechanisms helps us interpret the results.}. %

\begin{figure}[htp]
\begin{center}
\includegraphics[width=\columnwidth]{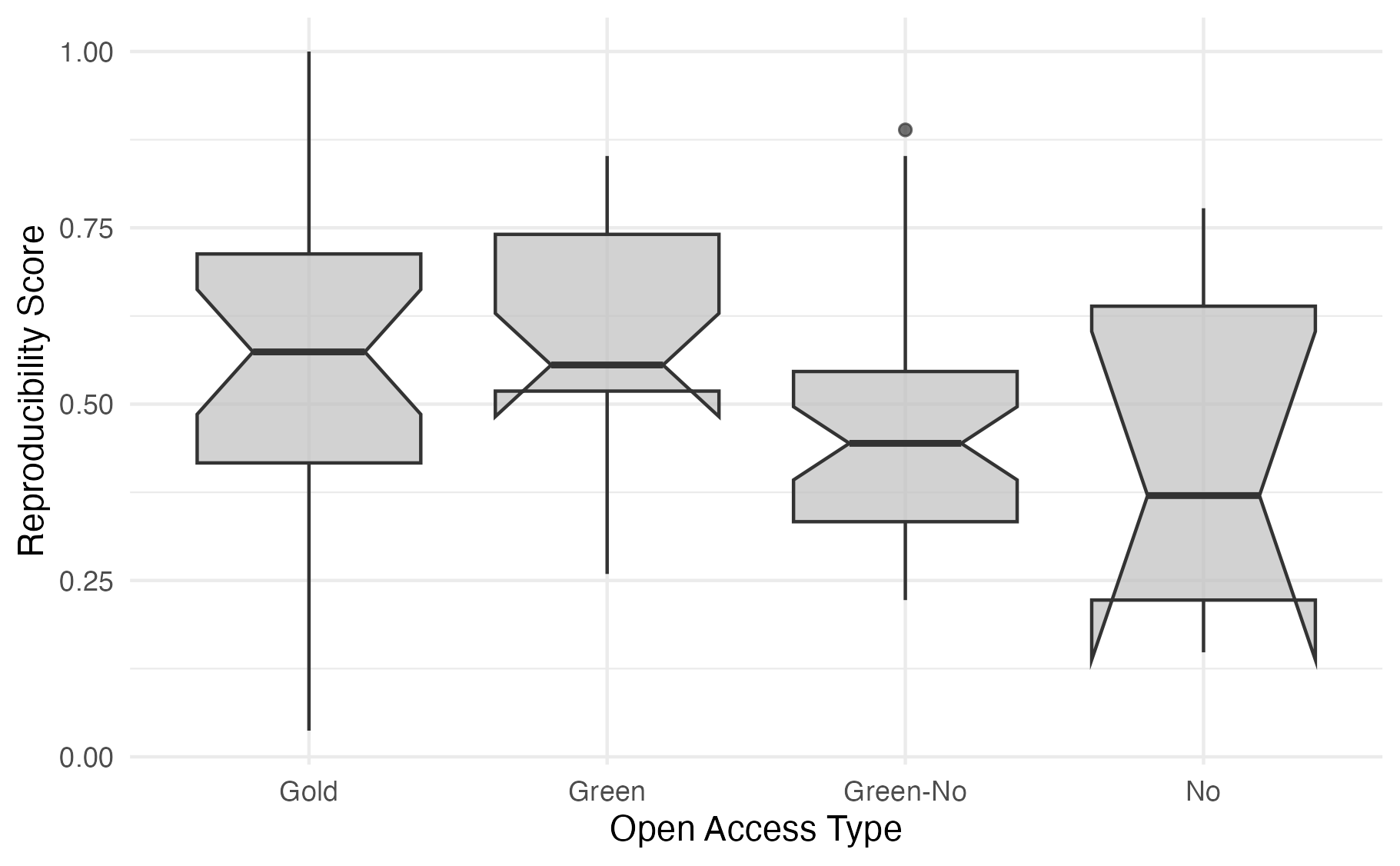}
\caption{Boxplots of Reproducibility vs Open Access Type}
\label{fig:OpenAccessBoxplots}
\end{center}
\end{figure}

Third, we examine the relationship between the type of Open Access for the paper (see Table~\ref{tab:AccessCts}) and Reproducibility. From the boxplots in  Fig.~\ref{fig:OpenAccessBoxplots} it would suggest that despite considerable variation within each category, there is some evidence that Gold and Green Access (where authors consciously choose to make their paper available) tend to be more reproducible than studies where the paper is protected by a paywall. Interestingly, the median score for Gold is lower than that for Green Access.  

\begin{figure}[htp]
\begin{center}
\includegraphics[width=\columnwidth]{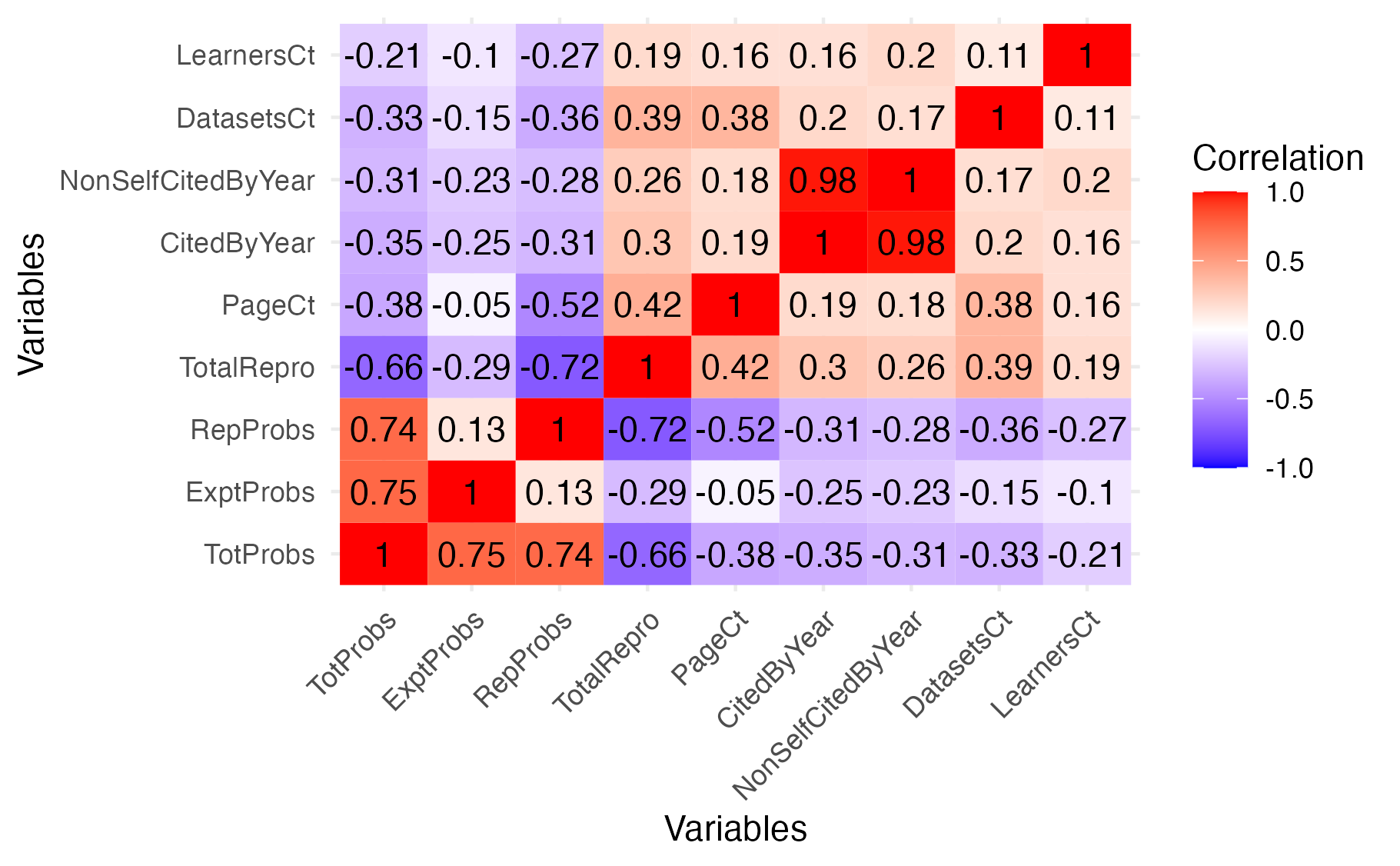}
\caption{Heat map of Spearman correlation coefficients}
\label{fig:SpearmanHeat}
\end{center}
\end{figure}

The relationship between reproducibility and citations per year is weak (Spearman's correlation coefficient = 0.24) indicating limited evidence that more reproducible work is more frequently cited. Nor is there any obvious relationship with experimental design such as the number of datasets (see the correlation matrix in Fig.~\ref{fig:SpearmanHeat}.  Note also, the stronger correlations are the consequence of functional relationships, so Total Problems = Experimental Probs + Reporting Problems.  Finally, as one might expect, reporting problems are negatively correlated (-0.7) with Total Reproducibility. 

Overall, we see a wide range of reproducibility scores with an unimpressive median score of 0.52 (or 14/27 questions obtaining positive scores). There is some evidence to suggest that journal papers tend to be more reproducible than conference papers. Other patterns such as those associated with page length and citations are less clear cut.

\subsection{RQ4: What kinds of quality issues are found in SDP studies?}

Here, we summarise the prevalence of the issues outlined in Section~\ref{Sec:Method} that are demonstrably weaknesses or problems in (i) the design and execution and (ii) the reporting of a computational experiment. In undertaking this audit we have sought to consider the context of an experiment; so, for example, if a study does not make use of statistical significance testing then it is not relevant to consider whether significance (alpha) thresholds need adjusting. For some studies, time ordering imposes a single analysis framework and hence there is only a single set of results and reporting the spread or dispersion is not relevant. Taking this into account, we tabulate our overall findings by problem (in decreasing order of prevalence) in Table~\ref{Tab:ProbCts}.  

\begin{table}
\centering
\caption{Frequencies of issues in studies}
\label{Tab:ProbCts}
\begin{tabular}{lr}
\hline
Issue                                       & Count  \\
\hline
Using problematic metrics                     & 68     \\
Not addressing spread as well as location     & 59     \\
Benchmarks and better than random             & 43     \\
Lack of OOS-validation                        & 31     \\
Multiple statistical tests without correction & 18     \\
\hline
Total study design / implementation issues & 219 \\
\hline
No link to code                               & 82 \\
Behind a `paywall'                            & 50 \\
Low reproducibility                           & 42 \\
No link to data                               & 34 \\
\hline
Total issues in reporting                     & 208 \\
\hline
Grand Total                                   & 427 \\   
\hline
\end{tabular}
\end{table}

\begin{figure}[htp]
\begin{center}
\includegraphics[width=\columnwidth]{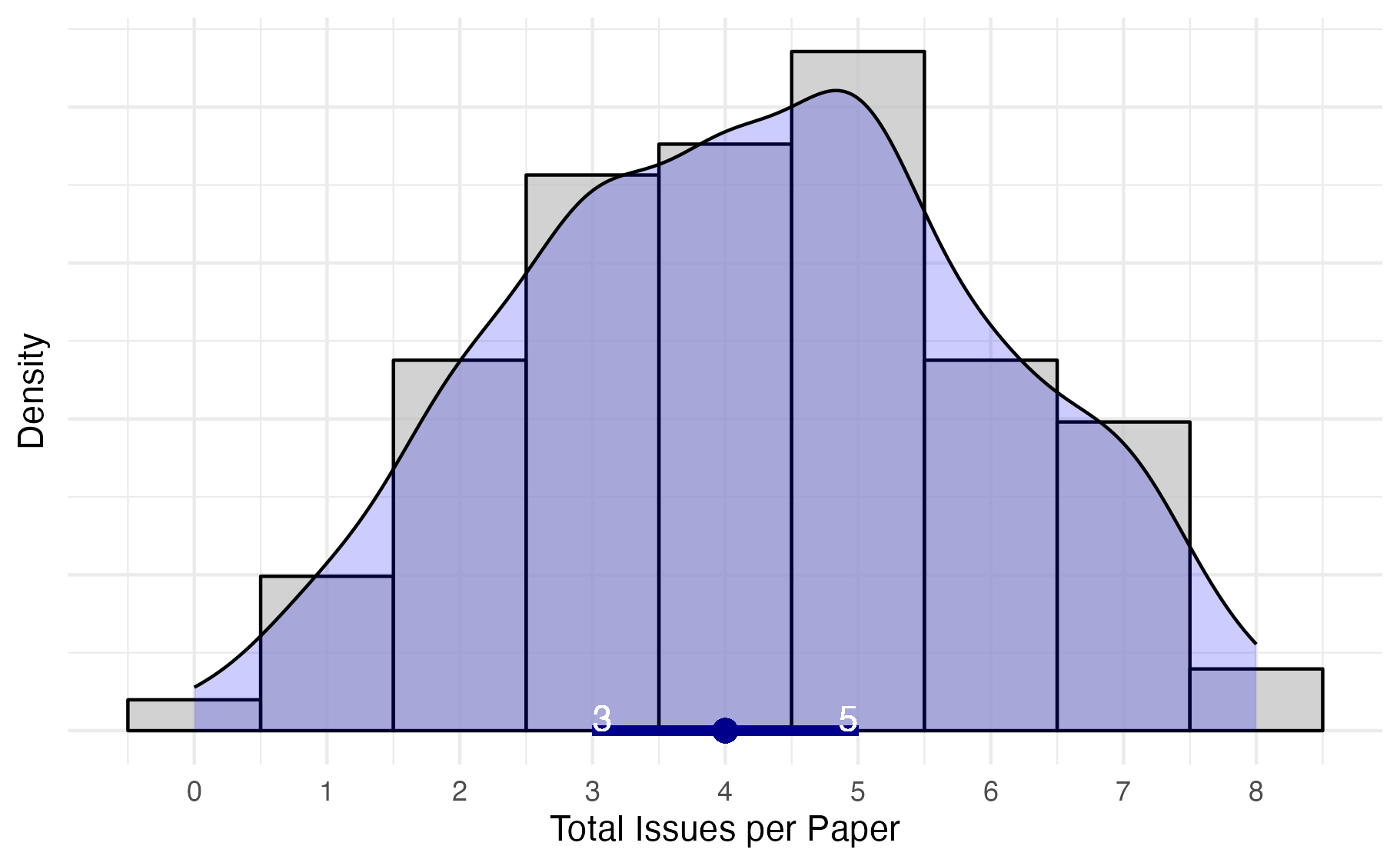}
\caption{The distribution of total issues per paper}
\label{fig:ProbsHist}
\end{center}
\end{figure}

Next, we address how issues distribute by individual paper. In total, we found 427 issues distributed across 101 papers (see Fig.~\ref{fig:ProbsHist}).  These ranged from 0 to 8 with a median of 4 and an IQR of 3-5. While not all the issues will invalidate the findings of a paper, e.g., placing a paper behind a `paywall' does not mean the analysis is incorrect. However, even this hinders independent scrutiny and neither serves the research community nor Open Science well. Notably, only a single paper had zero issues.

\begin{figure}[htp]
\begin{center}
\includegraphics[width=\columnwidth]{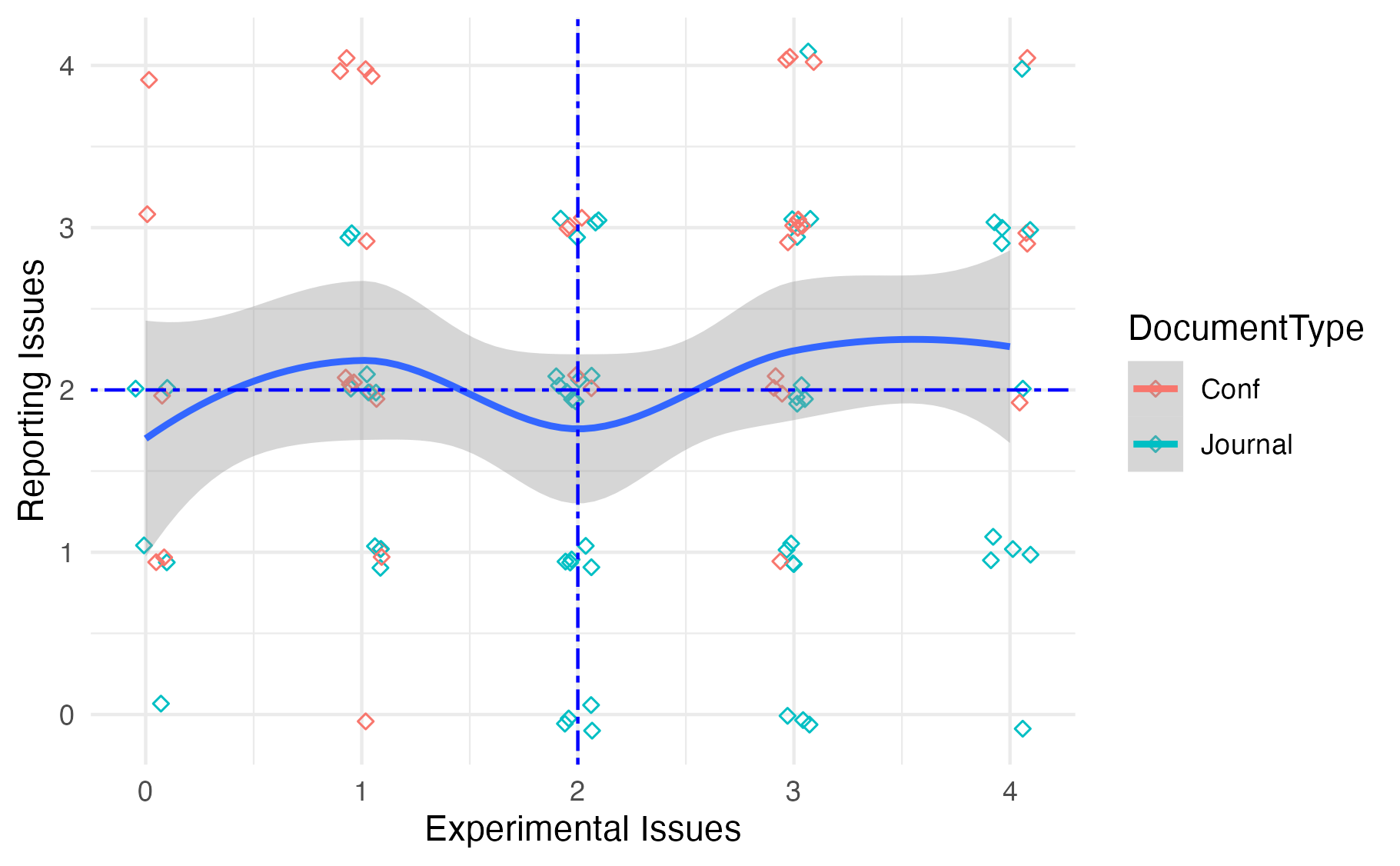}
\caption{Experimental versus reporting issues per paper}
\label{fig:ExptVreportingIssuesPlot}
\end{center}
\end{figure}

The scatterplot (see Fig.~\ref{fig:ExptVreportingIssuesPlot}) shows the distribution of types of issue (experimental and reporting) by paper. It is clear that they are essentially independent. The blue dashed lines represent the medians.

\begin{figure}[htp]
\begin{center}
\includegraphics[width=\columnwidth]{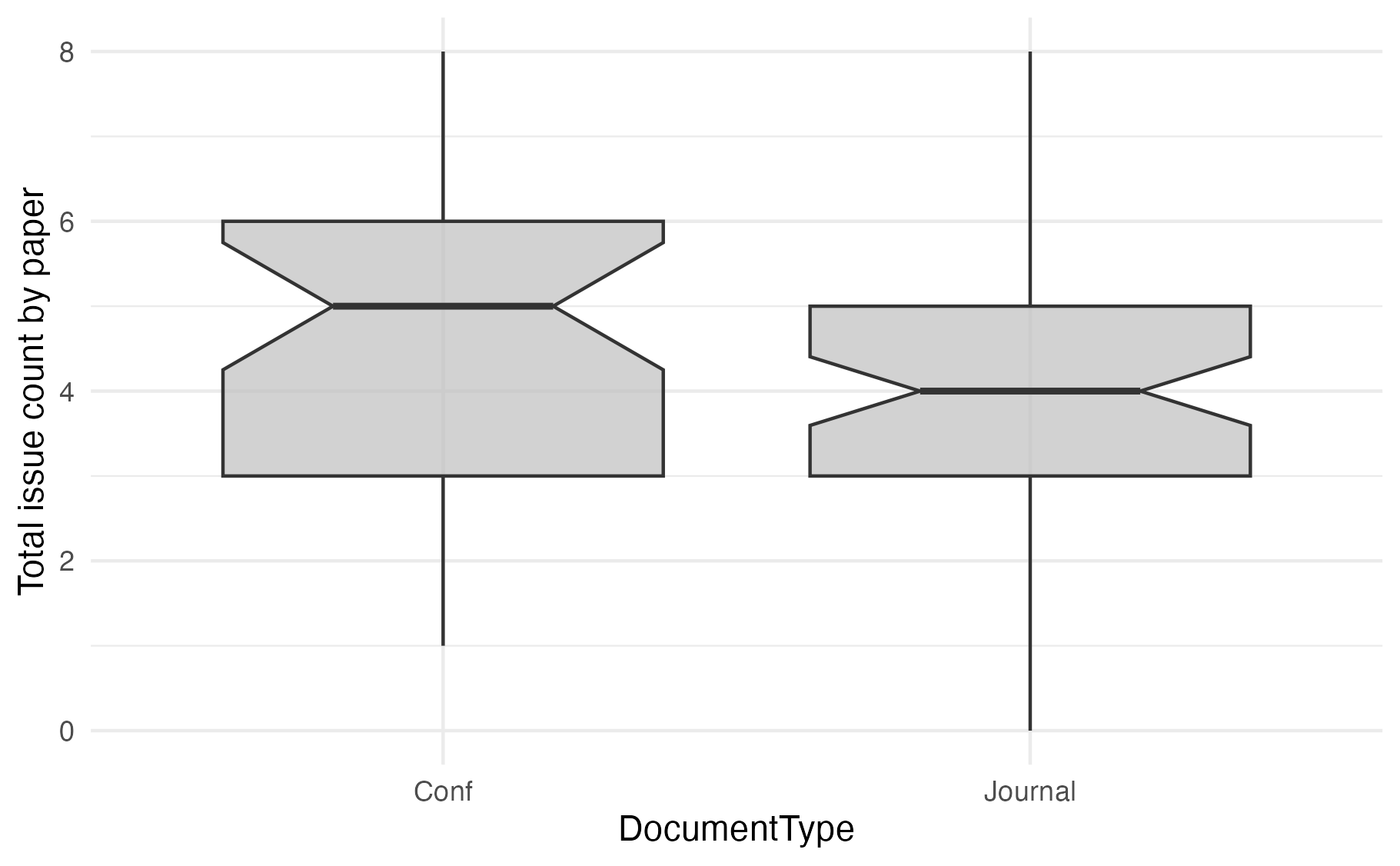}
\caption{Boxplots of issue counts of journal and conference papers}
\label{fig:ProbCtDocTypeBoxplots}
\end{center}
\end{figure}

Second, we explore whether there is a difference between conference and journal papers for issues.  Examining the boxplots in Fig.~\ref{fig:ProbCtDocTypeBoxplots} would suggest that there is some tendency for journal papers to contain fewer issues than conference papers with medians of 4 and 5 respectively.  The notches, representing the 95\% confidence intervals, do not overlap, suggesting the difference is non-trivial.  

\begin{figure}[htp]
\begin{center}
\includegraphics[width=\columnwidth]{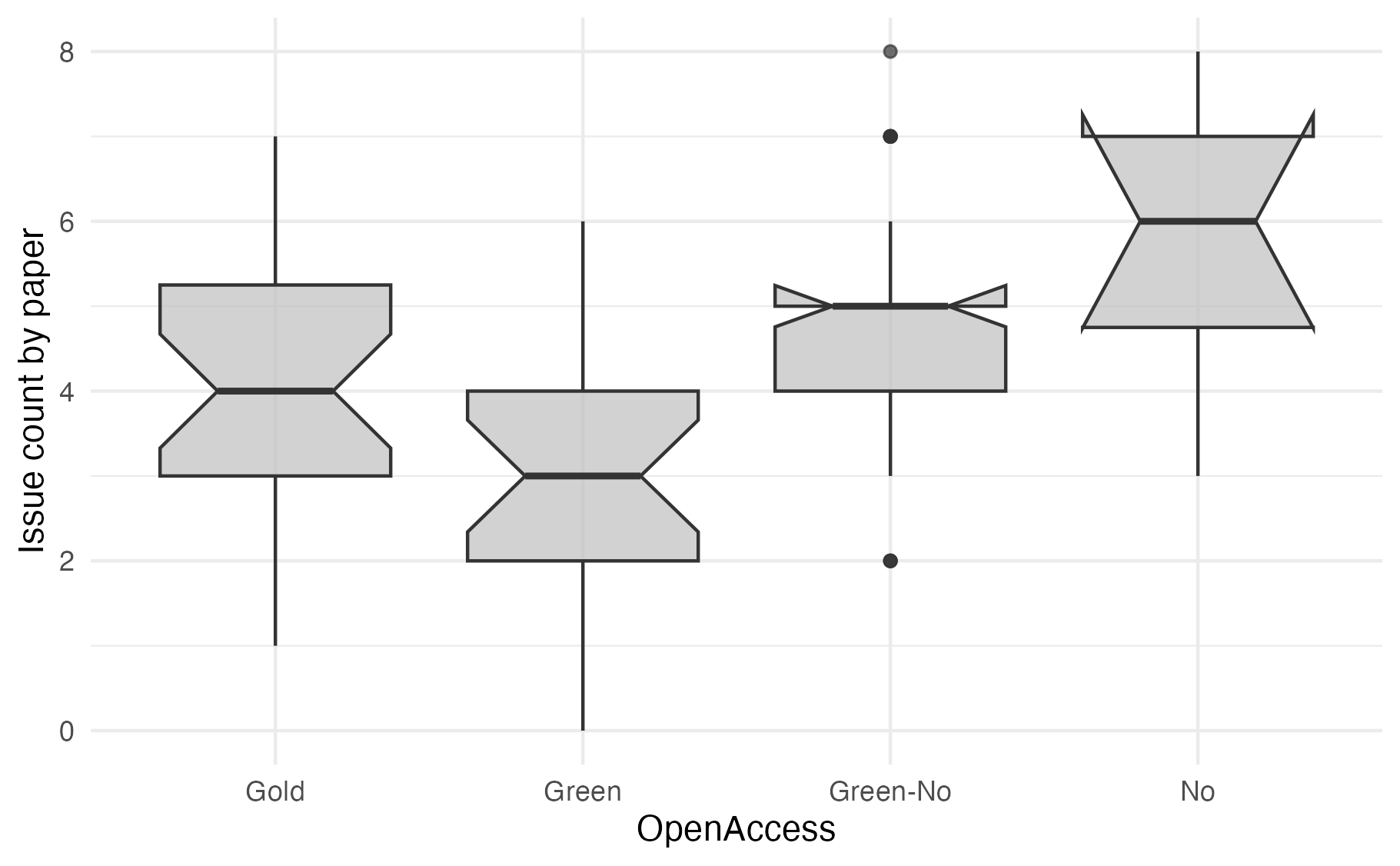}
\caption{Issues per paper by open access type}
\label{fig:ProbCtOABoxplots}
\end{center}
\end{figure}

A different approach is to think of the effect size as the difference between the medians for conference and journal papers. Using a bootstrap to establish the 95\% confidence intervals for the effect, we see the difference is: 95\% CI: 0.037, 0.222 which does not cover a zero or negative effect. Hence, again this gives some support for a \textit{small} difference in issue counts between journal and conference papers.  Note that this effect disappears when we \textit{only} consider experimental issues (see Fig.~\ref{fig:ExptVreportingIssuesPlot}) where the scatter of journal and conference data points shows little pattern on the x-axis for experimental issues.
We also noted that the open papers tended to have fewer issues (see the side by side boxplots in Fig.~\ref{fig:ProbCtOABoxplots} where the CI notches of the open papers do not overlap the paywalled papers).  It is not clear whether this is a causal relationship, but making a research paper more widely available does not do harm and might possibly encourage authors to be more methodical.

\begin{figure}[htp]
\begin{center}
\includegraphics[width=\columnwidth]{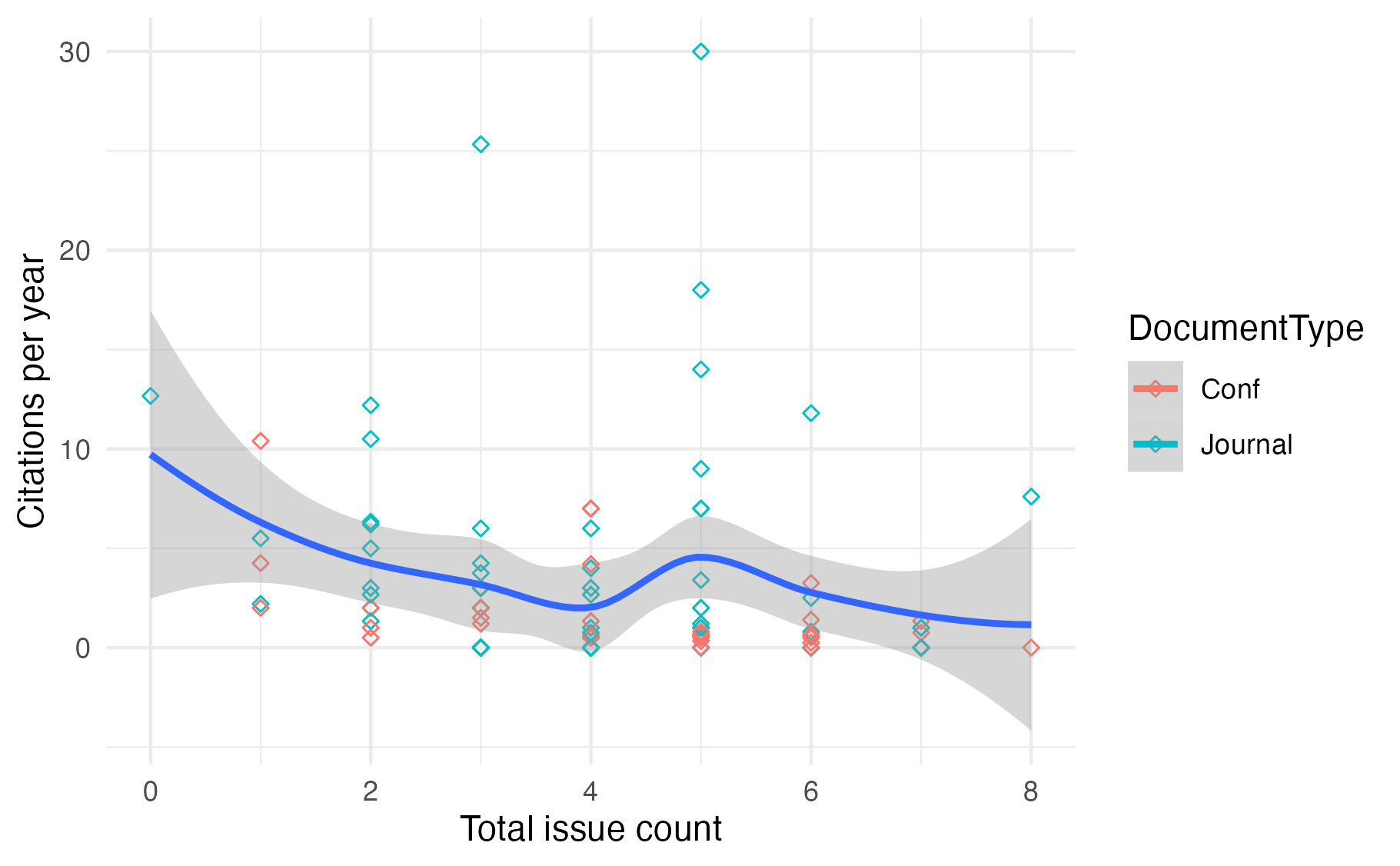}
\caption{Issues and citation counts}
\label{fig:IssuesVsCitations}
\end{center}
\end{figure}

Again, we checked for any association with normalised citation counts (see Fig.~\ref{fig:IssuesVsCitations}). The Spearman correlation is -0.21 indicating a weak negative association; however, inspection of the scatterplot suggests much deviation and the fitted loess smoother is flat for much of the range of issue counts.

\begin{figure}[htp]
\begin{center}
\includegraphics[width=\columnwidth]{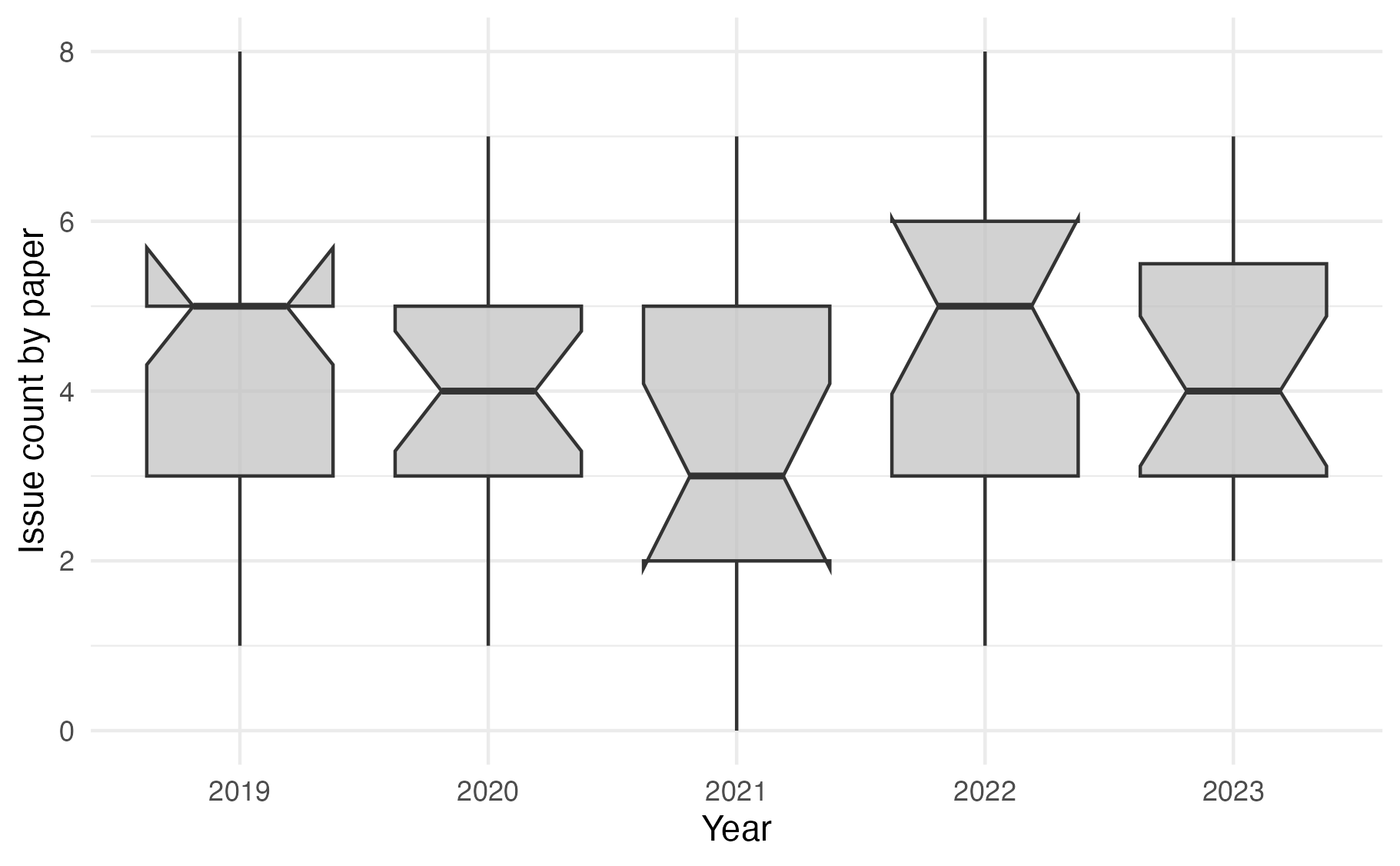}
\caption{Issues per paper by year}
\label{fig:ProbCtYearBoxplots}
\end{center}
\end{figure}

Finally, we consider whether there are any trends over time in Fig.~\ref{fig:ProbCtYearBoxplots} which reveals that, despite the year by year variability, there is no evidence to suggest that there is an overall effect over time. Unfortunately, we do not seem to be getting better, at least over the the five year window of this audit.

To summarise, the prevalence of issues with papers are widely distributed (almost all papers were impacted to some extent) and do not seem to be improving over time. Typically, a paper has four detected issues. There is weak evidence to suggest a relationship with Open Access publishing. Citation behaviour seems unrelated to either reproducibility or the count of issues.

\section{Discussion}\label{Sec:Discussion}

Our findings can be summarised as follows.
\begin{enumerate}
    \item There is great diversity in the design of experiments including the number of learners, datasets and performance metrics. Scientific heterodoxy can be a powerful means to advance science but it also complicates reporting, understanding and making fair comparisons.
    \item There is far less diversity in the choice of datasets (if not the absolute number), so a few curated collections such as the Promise datasets predominate. This means there is less independence than might be assumed for secondary analyses and meta-analysis. This can lead to difficulties with over-fitting to a repository curated more for reasons of convenience than representativeness.  Ironically, as Hand points out \cite{Hand2006}, the more successful a repository becomes, the greater the threat of distortion. It also means that the standard assumptions regarding random samples from some defined population are somewhat challenging to maintain.
    \item Approximately 40\% of the experiments explicitly undertake some kind of strategy to deal with imbalanced training data (where the prevalence of the positive case i.e., containing defects is low). Given the ubiquity of imbalanced datasets, this seems surprisingly low and will leave the analysis of results vulnerable to the potential biases of some of the performance metrics.
    \item Performance metrics that are widely accepted as being problematic --- certainly in the context of two-class problems and imbalanced datasets --- continue to be deployed, with more than 60\% of studies using F1 and 35\% of studies using Accuracy. This can substantially impact results and the preference ordering of the learners under evaluation \cite{Powers2011,Yao2021}. 
    \item Almost half of the experiments do not use any chance-anchored metric, so it is not possible to determine whether an algorithm is actually performing better than guessing. This is problematic given that other studies, e.g.,~\cite{Yao2020} have found a non-trivial number of learners doing worse than random.
    \item Again, there is much diversity in the choice of validation strategies to simulate out-of-sample prediction which is problematic because the choice of validation strategy is well known to impact results \cite{Kohavi1995,Stapor2021}. Clearly, this is an important area. The most common choice is 10-fold cross-validation, but this may be less relevant to how the SDP would actually be deployed than those based on cross-project and JIT prediction. More worrying is that some studies \textit{appear} not to use any method so presumably are model-fitting; other papers are simply unclear on the matter.
    \item Statistical errors affect almost a quarter of experiments where the acceptance or significance threshold is not adjusted for undertaking many, often in excess of 100, inferential tests.  While we are of the view that estimating effect size is more valuable than significance testing, for this audit we adopt an agnostic stance. Self-evidently, acceptance thresholds, where used, must be adjusted in the face of multiple tests to prevent unacceptable false positive rates \cite{Shaffer1995}. Approaches include Benjamini-Hochberg \cite{Benjamini1995} and Nemenyi post hoc testing \cite{Demsar2006}. To ignore this strikes us as quite problematic in that inferences and conclusions are directly impacted by the inflated possibility of false-positives.
    \item Spread needs reporting as well centre. This may appear trivial, however, it is reasonable to communicate the variability of predictive results as well as typical values such as the mean. From a practitioner point of view, understanding the range of possible prediction outcomes can be important.
    \item Informally, we found that for at least some of the papers, the descriptions of experiments were both hard to follow and omitted relevant information.  More formally, our use of the Gonz{\'a}lez-Barahona and Robles instrument for assessing reproducibility revealed a good deal of scope for improvement. There may be occasions when an experiment is designed better than we judged, but without the relevant details we cannot know this. Replication and meta-analysis are also hindered.
    \item Surprisingly few (50.5\%), to us at least, of the studies were shared as open access. There is also some weak association between open access and reproducibility such that Gold and Green Open Access papers seemed to be more reproducible.
    \item Journal papers tend to be more reproducible than conference papers although whether this due to typically being allowed more pages or a more constructive and iterative refereeing process is unclear.
    \item Overall, we identified 427 issues with only one paper being entirely issue-free.  The median number of issues is four. These split into 219 experimental issues and 208 reporting issues. The medians for both these are two.
    \item Citation counts, even when taking into account the paper age do not seem to be a guide as to paper quality (either reproducibility or problem count). Given high citation counts are widely taken to connote high research quality or importance, this is both unexpected and a little concerning.
\end{enumerate}

So what practical lessons can be derived?  These can be organised into three groups: (i) relating to the design of the computational experiment, (ii) to the reporting of the experiment and (iii) for the wider research community.

\begin{description}
    \item [\textbf{Experimental design recommendations:}] for the design and conduct of computational experiments.
\begin{enumerate}[label=R\arabic*] 
    \item It is important to explicitly deploy some form of validation procedure to simulate prediction of out-of-sample instances and prevent over-fitting.  We suggest that stochastic procedures such as k-fold cross-validation need $k \ge 10$ folds, or where large sample conditions do not hold then more repetitions, e.g., $2 \times 5$ folds \cite{Kohavi1995,Wong2020}.  However, the closer the procedure is to real-world usage the more useful the experimental results.  Therefore, respecting time ordering, cross-project and JIT validation are important considerations.
    \item Use a chance-anchored performance metric, e.g., AUC or MCC or additionally report a simple metric such as Bookmaker's Odds \cite{Powers2011}.  Without this, it can be unclear whether a learning algorithm is doing better than guessing. With unbalanced datasets, values from metrics such as F1 can be quite unintuitive to interpret.
    \item Do not use problematic metrics: this seems self-evident and is widely discussed (e.g., \cite{Powers2011,Chicco2020,Yao2021}) and easy to implement.
    \item Do not focus solely upon the prediction results central tendency but report dispersion. This may appear trivial, however, it is important to communicate the variability of predictive results as well as typical values such as the mean. A five-number summary or boxplots are examples of appropriate solutions. Many learning algorithms and OOS-validation strategies are stochastic, so understanding the variability of results is important especially to practitioners. 
    \item If using multiple statistical inference tests, control for the false discovery rate. We recommend a modern approach such as proposed by Benjamini-Hochberg \cite{Benjamini1995} which is less conservative than traditional approaches such as the Bonferroni correction.
    \item In general, unless there are stronger methodological reasons to the contrary, restrict freedom to produce ever more complex and individual experimental designs, since this makes experiments harder to analyse, understand and comparisons between studies more awkward.
\end{enumerate}
\item[\textbf{Reporting recommendations:}] that relate to the reporting of both the experimental method and results.
\begin{enumerate}[label=R\arabic*, resume]
    \item Fully report the experiment to include data (this is usually done), details of data pre-processing and analysis code (less commonplace). This should include all confusion matrices to allow alternative metrics to be computed, if desired.
    \item Where relevant, be clear what version of any dataset is used.
    \item Carefully describe all data cleaning or pre-processing to enable work to be reproduced.
    \item Use and report seeds for any stochastic analysis, e.g., the random allocation of instances to folds.
    \item Consider the durability of web links. We found that approximately 35\% of links were broken in just the five year period covered by the audit. 
    \item Behind a `paywall' is related to reproducibility, since if many scientists are prevented from accessing a paper then they cannot build upon it. Therefore it is important to ensure an open-access version of a paper is available.  For most publishers, this simply means posting the final postprint on a not-for-profit site such as arXiv or the author's own institution.
\end{enumerate}

\item[\textbf{Community recommendations:}] here we consider some recommendations for researchers and the wider community.
\begin{enumerate}[label=R\arabic*, resume]
\item We need careful reading of papers by reviewers, editors, readers and citers of published papers. Some of the issues we have detected, in particular the absence of explanation and use of tortured phrases, should be visible even upon a cursory reading. It would also be valuable to make greater use of public reviewing platforms such as PubPeer.
\item Expect sharing of code and results, as well as data. The idea of data sharing is widely accepted, however sharing of code and raw results far less so.  Improving upon this would help both better detection of issues and easier reproducibility.
\item There may be value in the community providing more guidance and even prescription for the design of experiments. It is arguable if researchers have too many degrees of freedom this makes errors easier to commit and the research harder to understand.
\item Journals seem to offer advantages for level of detail required for reproducibility.  Here, we venture a little beyond our audit but there do seem benefits to the journal publication model as compared to conferences. Of course, conferences serve many purposes over and above publication vehicles. However, if software engineering were to place higher value upon journal papers we see it as a good thing.
\end{enumerate}
\end{description}

\section{Threats to Validity}\label{Sec:Threats}

\subsection{Scope of Audit Checks}

One potential limitation is the limited scope of checks we have performed during our audit. While our analysis is wide-ranging, it does not cover all possible types of errors or inconsistencies that could be present in the studies. For example, we did not scrutinise the internal consistency of the reported confusion matrices, an aspect highlighted in prior research as a source of error \cite{Shepperd2019}.

\subsection{Measurement Error}

Our approach to assessing reproducibility could itself be subject to measurement error. The metrics we used to evaluate reproducibility may not fully capture the nuances and complexities involved in replicating a study.  In following the Gonz{\'a}lez-Barahona and Robles~\cite{Gonz2012} instrument, this does mean that each reproducibility item is equally weighted which is a simplifying assumption.  However, we are of the view that gross differences between reproducibility levels will be highlighted even if small differences need to be treated with more circumspection.

\subsection{Rater Bias and Subjectivity}

The audit involved subjective judgments, such as the categorisation of different research designs. Despite efforts to standardize these evaluations, the possibility of rater bias cannot be entirely ruled out. To mitigate this, we used one paper for initial training and also independently double-checked and agreed all remaining studies.  

\subsection{Sample Size and Representativeness}

Our study included 101 papers, which, while substantial, is not exhaustive and we estimate represents of the order of 6.5\% of all the primary studies published 2019 to 2023. The papers were also selected based on specific inclusion criteria, possibly limiting the generalisability of our findings. The sample may not be fully representative of all work in the field, particularly studies published in venues not indexed by Scopus. However, if we restrict our population to studies that have been published in well-regarded venues and that have undergone meaningful peer review, then we have a reliable sample frame (the Scopus search results) and a demonstrably random sampling mechanism. Thus we can compute the standard error of sample statistics, e.g., the mean reproducibility score where the distribution is approximately symmetric. The sample Mean is 0.520, the adjusted (for the finite population correction) standard error of the mean 0.020 and the  
95\% Confidence Interval is [0.480, 0.560] which is reasonably precise i.e., $\pm 4\%$. 

We also considered the sample representativeness in terms of the \textit{perceived} quality of a venue.  Our sample includes 4 papers from \textit{IEEE TSE} and \textit{Information \& Software Technology} and 2 from \textit{Empirical Software Engineering}. By its nature a sample will not cover every venue but we have good reasons to believe both `prestigious' and `less prestigious' venues are adequately covered.

\subsection{Recommendations for Further Investigation}\label{SubSec:Rec}
To mitigate the threats to validity identified in this study, we suggest researchers should aim for a more exhaustive range of quality checks, possibly incorporating automated tools to scrutinise aspects like confusion matrices \cite{Bowes2014}. Future studies might also consider employing multiple metrics for assessing reproducibility to capture its various dimensions. Finally, to enhance representativeness of the results, future audits could aim for a larger sample sizes and consider including papers from other sources, not just those indexed by Scopus so that it might be possible to compare the grey literature with the more regular refereed literature covered by Scopus.

\section{Conclusion}\label{Sec:Concl}

To recap, in response to concerns about the quality and reproducibility of software defect prediction studies, we undertook an audit of experiments published from 2019-2023.  We sampled 101 experiments and reviewed each one for the choice of training data, cleaning strategies relating to imbalance, the validation strategy, performance metrics, statistical inference and reproducibility. Unfortunately, this identified quite widespread issues consistent with other previous studies such as \cite{Liem2020,Liu2021,Shepperd2019}. 

Our findings can be summarised as follows.
\begin{itemize}
\item There is great diversity in the design of experiments but less so in the choice of datasets, where the Promise datasets predominate.
\item Overall, we identified 427 issues (not all fatal, but minimally, unnecessary and unhelpful) with only one paper being entirely issue-free. The median number of issues was four per paper and cover both experimental design and reporting. Some of the more prestigious outlets, for example Transaction type papers are not immune.  It should also be stressed our search for research quality issues is by no means exhaustive since we focused on issues that are easy to detect and cannot be considered a matter of opinion.
\item Citation counts, even when normalised, do not seem to be a guide as to paper quality (for either reproducibility or issues).
\item Journal papers tend to be more reproducible and contain fewer issues than conference papers although whether this is due to typically being allowed more pages or a more constructive and iterative refereeing process is unclear.
\item We found 2 out of 101 papers to contain clear examples of tortured phrases, e.g., “software insects”, which could be indicative of paper mill activity.
\end{itemize}

Second, we list a total of 16 lessons that emerge from our audit to those conducting computational experiments to evaluate software defect prediction. These are relevant to researchers, reviewers and editors and to the wider software engineering community. One source of encouragement is that none of them are controversial or difficult to implement.

Finally, we wish to be clear that scientists are human.  Furthermore, computational experiments, by their very nature tend to be complex. We all --- the authors of this paper included --- make mistakes, thus we do not wish this audit to be seen as some kind of schadenfreude-piece. Nevertheless, if we consider the astonishing amount of effort deployed in researching software defect prediction (considerably more than 1,500 primary studies in the past five years alone) then the lack of reliability, insight or impact should make us wish to do better.

\section*{Data and Code Availability Statement}
The dataset used in this study and the code to analyse it are available from here \url{https://zenodo.org/records/17696089} and DOI \url{https://doi.org/10.5281/zenodo.13927601}.

\section*{Conflict of Interest}
The authors declared that they have no conflict of interest.

\section*{Funding}
No specific funding was received for conducting this study.

\section*{Author contributions}
All authors contributed to the study conception, design. Material preparation was performed by Martin Shepperd. Data collection and validation were performed by all authors. The data analysis was performed by Martin Shepperd and Leila Yousefi. The first draft of the manuscript was written by Giuseppe Destefanis, Martin Shepperd and Steve Counsell.  All authors commented on previous versions of the manuscript and all authors read and approved the final manuscript.

The order of authors was determined randomly.

\bibliographystyle{abbrv}
\bibliography{ML_Audit}

\section{Appendix: Raw data description}\label{sec:Appendix}

\begin{longtable}{|l|c|p{8cm}|}
  \hline
  Variable Name & Type & Comment \\ 
  \hline
  \endfirsthead

  \hline
  Column & Type & Comment \\ 
  \hline
  \endhead
  \hline
PaperID & character & Unique paper identifier that we can map back to the actual paper. \\ 
  Checker & factor &  Person extracting data.\\ 
  Checker2 & character & Person extracting data.  \\ 
  PredictionType & factor & Either classification or regression.\\ 
  Year & factor &  Year of publication (taken from Scopus).\\ 
  \hline
  SourceTitle & character & Venue published e.g., journal or conference name. \\ 
  Core\_A & factor & Publication venue is rated A*/A by CORE (Y or N). \\
  PageCt & integer &  Length of article (taken from Scopus).\\ 
  CitedBy & integer &  Citation count (taken from Scopus).\\ 
  SelfCited & integer &  Self citations by any co-author.\\ 
  OpenAccess & factor &  Is the paper publicly available (Y or N)?\\ 
  \hline
  DocumentType & factor & Conf(erence) or Journal.\\ 
  Tortured & factor & Does the paper contain clear examples of ``tortured phrases'' (Y or N)? \\ 
  Summary & character &  Free format narrative description by Checkers\\ 
  Datasets & character & Free format description of the datasets or repositories used.\\ 
  DatasetsCt & integer & How many distinct datasets  used?  Different releases are separate data sets eg Camel 1.0, 1.2 = 2 data sets.  \\ 
  \hline
  LearnersCt & integer &  The number of learners/algorithms/treatments being evaluated.  This might include pre-processing such as transformation or feature subset selection. Usually obvious from the results table(s).\\ 
  MetricsText & character & What classification performance metrics are collected? Items separated by comma. \\ 
  F1 & integer &  F1 is often referred to as the F-measure or even F-score (1 if used).\\ 
  MCC & integer & Matthews correlation coefficient (1 if used). \\ 
  AUC & integer & Area under the curve (AUC).  Sometimes called ROC (1 if used). \\ 
  \hline
  Accuracy & integer & (1 if used). \\ 
  Precision & integer &  Is TP/(TP+FP) (1 if used).\\ 
  Recall & integer &  Also known as sensitivity or TPR ie TP/(TP+FN) (1 if used).\\ 
  Specificity & integer & Also known as TNR ie TN/TN+FP (1 if used).\\ 
  Other & integer & The count of other metrics so if non-zero the total in the next column is correct. \\
  \hline
  MetricsCt & integer & Total count of different performance metrics used in the analysis. \\ 
  SummaryText & character &  Free format text on how the metrics summarised e.g., from multiple folds or multiple data sets?  (Mean or boxplots are common).\\ 
  Location & factor & Is location reported e.g., mean or median  (Y or N)?\\ 
  Spread & factor & Is spread or dispersion reported, e.g., IQR or variance (Y or N)?\\ 
  RandomBenchmark & factor & Do the authors compare prediction performance with chance e.g., Bookmakers odds or a chance-anchored metric such as AUC? (Y or N)?\\ 
  \hline
  StatInference & factor & Do the authors make use of statistical inference, most likely via p-values and some significance threshold? (Y or N)?\\ 
  TestsCt & integer & How many statistical tests are reported (typically in tables of dataset v learning algorithm and then for each metric).  Report NA if no tests (Y or N or NA)?\\\ 
  ImbalancedMethod & factor & E.g., under/over sampling, SMOTE etc., but NA where irrelevant e.g., regression. (Y or N or NA)?\\ 
  AdjustSig & factor & Do the authors adjust significance threshold for multiple tests, otherwise NA if 0 or 1 test (Y or N or NA)?\ \\ 
  OutOfSample & factor & Is CV or other out of sample validation method e.g., bootstrap used or not and NA if CV isn't applicable to the study. (Y or N or NA)?\\ 
  \hline
  TypeCV & character & Free format description of CV method. \\ 
  m & integer & Number of replications of the validation process or NA.\\ 
  n & integer & Number of folds for CV or NA. \\ 
  \hline
   &  &  The next 27 columns address the \textbf{reproducibility} of a study as per González-Barahona \& Robles 2012.  0 = not present/not ok; 1 = present / ok. \\
\hline
 RawIdentification & integer &  Is it clear where the (original) raw data be obtained from?\\ 
  RawDescription & integer &  Is the published information about the raw data, including its internal organization and structure and its semantics sufficiently detailed?\\ 
  RawAvailability & integer & Is it easy for a researcher to obtain the raw data, or have access to it? \\ 
  RawPersistence & integer & Is it likely the raw data will be available in the future? \\ 
  RawFlexibility & integer &  Is the raw data flexible, how easily can it be adapted to new environments?\\ 
  \hline
  ExtractionIdentification & integer &  Is it clear where the (original) extraction method be obtained from? \\ 
  ExtractionDescription & integer &  Is the published information about the extraction method sufficiently detailed?\\ 
  ExtractionAvailability & integer & Is it easy for a researcher to obtain the extraction method, or have access to it? \\ 
  ExtractionPersistence & integer & Is it likely the extraction method will be available in the future? \\ 
  ExtractionFlexibility & integer &  Is the extraction method flexible, how easily can it be adapted to new environments?\\ 
  \hline
  ProcessedIdentification & integer &  Is it clear where the processed dataset be obtained from?\\ 
  ProcessedDescription & integer & Is the published information about the processed dataset sufficiently detailed? \\ 
  ProcessedAvailability & integer & Is it easy for a researcher to obtain the processed dataset, or have access to it?  \\ 
  ProcessedPersistence & integer &   Is it likely the processed dataset will be available in the future?\\ 
  ProcessedFlexibility & integer &  Is the processed dataset flexible, how easily can it be adapted to new environments?\\ 
  \hline
  AnalysisIdentification & integer &  Is it clear where the analysis methodology/tools can be obtained?\\ 
  AnalysisDescription & integer &  Is the published information about the analysis methodology/tools sufficiently detailed?\\ 
  AnalysisAvailability & integer &  Is it easy for a researcher to obtain the analysis methodology/tools, or have access to it? \\ 
  AnalysisPersistence & integer &  Is it likely the analysis methodology/tools will be available in the future?\\ 
  AnalysisFlexibility & integer & Are the analysis methodology/tools flexible, how easily can they be adapted to new environments? \\ 
  \hline

  ResultsIdentification & integer & Is it clear where the results dataset be obtained from? \\ 
  ResultsDescription & integer & Is the published information about the results dataset sufficiently detailed? \\ 
  ResultsAvailability & integer & Is it easy for a researcher to obtain the results dataset, or have access to it?  \\ 
  ResultsPersistence & integer &  Is it likely the results dataset will be available in the future? \\ 
  ResultsFlexibility & integer &  Is the results dataset flexible, how easily can it be adapted to new environments?\\ 
  \hline
  Identification & integer & Are all the study parameters clearly identified? \\ 
  Description & integer & Are all the study parameters adequately described including their setting(s)? \\ 
  TotalRepro & numeric &  The overall reproducibility score out of 27 and then normalised 0-1. \\ 
  \hline
  DataLink & factor &  Is there a link to the data used (Y or N)?\\ 
  CodeLink & factor &  Is there a link to the code to generate the analysis (Y or N)?\\ 
  BrokenLink & factor &  NA if no links, Y if one or more links are broken otherwise N. \\ 
  CitedByYear & numeric & Total citations divided by years available. \\ 
  NonSelfCitedByYear & numeric &  (Total citations minus self-citations) divided by years available. \\ 
  \hline
  CVprob & numeric &  Are there issues relating to the absence of cross-validation where expected (or any other appropriate out-of-sample testing)? \\ 
  Randprob & numeric & Is there an issue due to the absence of random prediction benchmarks? \\ 
  BadMprob & numeric &  Problematic use of performance metrics i.e., F1 or Accuracy? \\ 
  AdjustProb & numeric & Issue arising from failure to adjust significance threshold (alpha) when undertaking multiple significance tests? \\ 
  NoVarProb & numeric & Issue arising from the failure to report the spread or variability of results. \\ 
\hline
  LowReproProb & numeric &  Issue arising from low reproducibility of the study, defined as the lowest tertile.\\ 
  NotOpenPubProb & numeric &  Issue arising from unavailability of the paper due to ``pay walls''.\\ 
  NotOpenDataProb & numeric & Issue arising from the unavailability of the data. \\ 
  NotOpenCodeProb & numeric &  Issue arising from not sharing analysis methods. \\ 
 \hline 
  TotProbs & numeric & Total count of issues (ranges from 0 to 9).\\ 
  ExptProbs & numeric & Total count of experimental design issues (ranges from 0 to 5). \\ 
  RepProbs & numeric &  Total count of reporting issues (ranges from 0 to 4). \\ 
   \hline
\end{longtable}

\end{document}